\documentclass[aps,prd,10pt,twocolumn,superscriptaddress,showpacs,longbibliography]{revtex4-1}

\pdfoutput=1
\usepackage{hyperref}
\usepackage{graphicx}
\usepackage{amsfonts,amsmath,amssymb,bm,bbm}
\usepackage{color}

\hypersetup{
    pdfnewwindow=true,    
    colorlinks=true,      
    linkcolor=blue,       
    citecolor=blue,       
    filecolor=blue,       
    urlcolor=blue         
}

\begin{document}

\title{TeV solar gamma rays from cosmic-ray interactions}

\author{Bei Zhou}
\email{zhou.1877@osu.edu}
\thanks{\scriptsize \!\! http://orcid.org/0000-0003-1600-8835}
\affiliation{Center for Cosmology and AstroParticle Physics (CCAPP), Ohio State University, Columbus, Ohio 43210, USA}
\affiliation{Department of Physics, Ohio State University, Columbus, Ohio 43210, USA}

\author{Kenny C. Y. Ng}
\email{chun-yu.ng@weizmann.ac.il}
\thanks{\scriptsize \!\! http://orcid.org/0000-0001-8016-2170}
\affiliation{Center for Cosmology and AstroParticle Physics (CCAPP), Ohio State University, Columbus, Ohio 43210, USA}
\affiliation{Department of Physics, Ohio State University, Columbus, Ohio 43210, USA}
\affiliation{Department of Particle Physics and Astrophysics, Weizmann Institute of Science, Rehovot 76100, Israel}

\author{John F. Beacom}
\email{beacom.7@osu.edu}
\thanks{\scriptsize \!\! http://orcid.org/0000-0002-0005-2631}
\affiliation{Center for Cosmology and AstroParticle Physics (CCAPP), Ohio State University, Columbus, Ohio 43210, USA}
\affiliation{Department of Physics, Ohio State University, Columbus, Ohio 43210, USA}
\affiliation{Department of Astronomy, Ohio State University, Columbus, Ohio 43210, USA}

\author{Annika H. G. Peter}
\email{apeter@physics.osu.edu}
\thanks{\scriptsize \!\! http://orcid.org/0000-0002-8040-6785}
\affiliation{Center for Cosmology and AstroParticle Physics (CCAPP), Ohio State University, Columbus, Ohio 43210, USA}
\affiliation{Department of Physics, Ohio State University, Columbus, Ohio 43210, USA}
\affiliation{Department of Astronomy, Ohio State University, Columbus, Ohio 43210, USA}

\date{1 August 2017}

\begin{abstract}
The Sun is a bright source of GeV gamma rays, due to cosmic rays interacting with solar matter and photons.  Key aspects of the underlying processes remain mysterious.  The emission in the TeV range, for which there are neither observational nor theoretical studies, could provide crucial clues.  The new experiments HAWC (running) and LHAASO (planned) can look at the Sun with unprecedented sensitivity.  In this paper, we predict the very high-energy (up to 1000 TeV) gamma-ray flux from the solar disk and halo, due to cosmic-ray hadrons and electrons ($e^++e^-$), respectively.  We neglect solar magnetic effects, which is valid at TeV energies; at lower energies, this gives a theoretical {\it lower} bound on the disk flux and a theoretical {\it upper} bound on the halo flux.  We show that the solar-halo gamma-ray flux allows the first test of the $\sim 5$--70 TeV cosmic-ray electron spectrum.  Further, we show that HAWC can immediately make an even stronger test with nondirectional observations of cosmic-ray electrons.  Together, these gamma-ray and electron studies will provide new insights about the local density of cosmic rays and their interactions with the Sun and its magnetic environment.  These studies will also be an important input to tests of new physics, including dark matter.
\end{abstract}

\maketitle


\section{Introduction}
\label{sec:introduction}
The Sun is a passive detector for cosmic rays in the inner Solar System, where direct measurements are limited.  It shines in gamma rays from its disk and from a diffuse halo~\cite{Seckel:1991ffa, Moskalenko:2006ta, Orlando:2006zs, Orlando:2008uk, Abdo:2011xn, Ng:2015gya}.  Disk emission is expected due to cosmic-ray hadrons interacting with solar matter, which produces pions and other secondaries of which the decays and interactions lead to gamma rays.  Halo emission is expected due to cosmic-ray electrons ($e^++e^-$) interacting with solar photons via inverse-Compton scattering.  There are no other important astrophysical mechanisms for steady solar gamma-ray production; solar-flare gamma rays are episodic, and are observed up to only a few GeV~\cite{Schneid1996a, Fermi-LAT:2013cla, Ackermann:2014rma, Pesce-Rollins:2015hpa}.

Gamma-ray observations thus open the possibility of detailed cosmic-ray measurements near the Sun. The hadronic and leptonic components can be distinguished because the disk and halo emission can be separated by direction.  Further, the energy spectra of the cosmic rays can be inferred from the gamma-ray spectra, which can be measured over a wide energy range.  This would give a significant advance compared to typical satellite detectors in the inner Solar System, which only measure the energy-integrated all-particle flux (e.g., Refs~\cite{JGRA:JGRA51485, JGRA:JGRA52795}), and are thus dominated by low-energy particles.  Further, gamma-ray data can trace the full solar cycle, testing how solar modulation of cosmic rays depends on energy and position~\cite{jokipii1971, Potgieter:2013pdj}.

\begin{figure}[!h]
\includegraphics[width=\columnwidth]{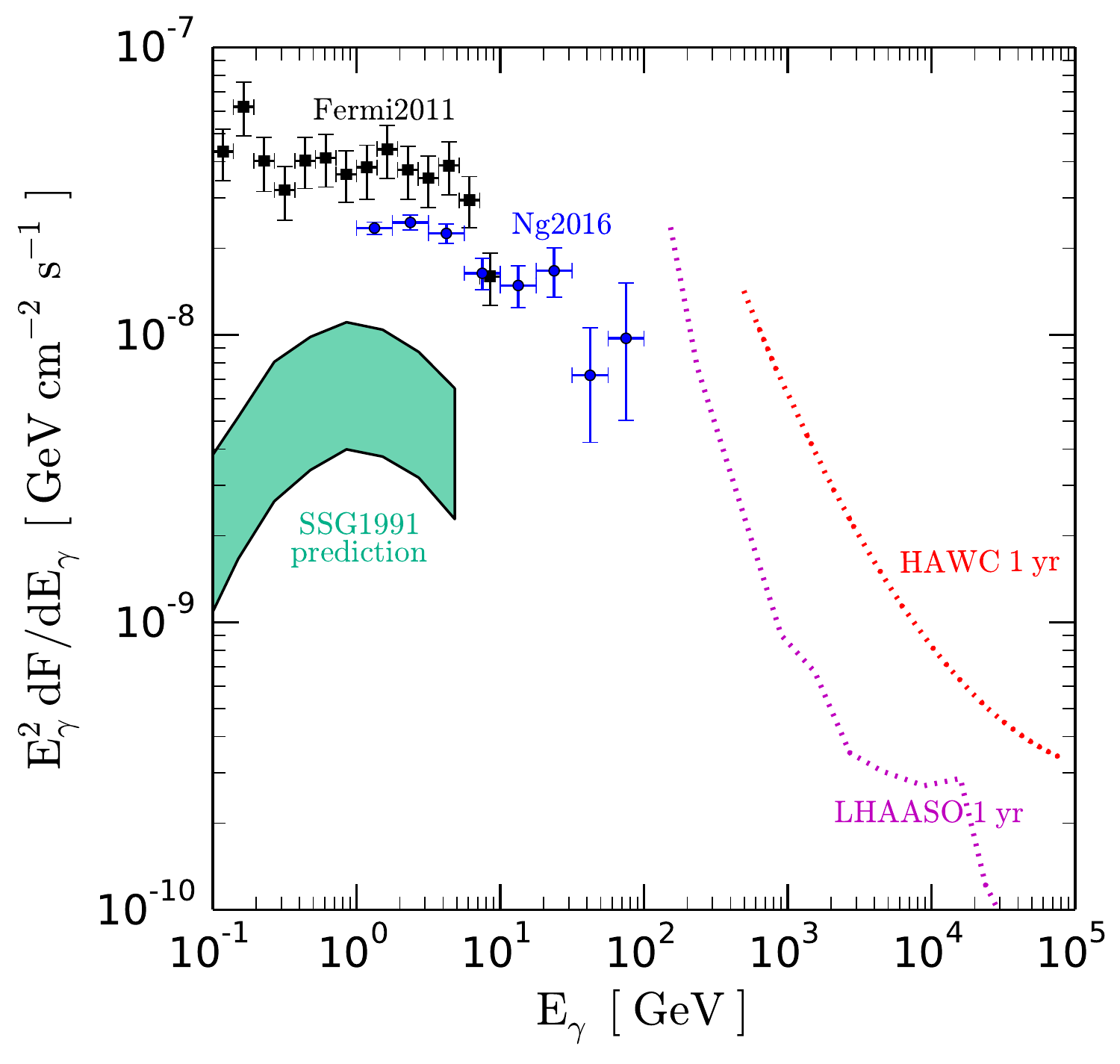}
\caption{Prospects for TeV solar gamma-ray observations, illustrated with the disk emission (details in Fig.~\ref{fig:detectability}).  Points: observations with Fermi~\cite{Abdo:2011xn, Ng:2015gya}, where the flux difference is due to time variation.  Green band: the only theoretical prediction that includes magnetic effects~\cite{Seckel:1991ffa}.  Dashed lines: the estimated differential point-source sensitivity of HAWC~\cite{Abeysekara:2017mjj} (scaled to one year) and LHAASO~\cite{Zhen:2014zpa, He:2016del}.
}
\label{fig:intro}
\end{figure}

Figure~\ref{fig:intro} shows that the prospects for measuring TeV solar gamma rays are promising.  The solar-disk fluxes measured in the GeV range with Fermi data~\cite{Abdo:2011xn, Ng:2015gya} are high, 
and naive extrapolation suggests that HAWC and LHAASO may detect gamma rays in the TeV range.  Further, the GeV observations are significantly higher than the theoretical prediction of Seckel {\it et al.}~\cite{Seckel:1991ffa}, who proposed a compelling mechanism by which the solar-disk gamma-ray flux could be enhanced by magnetic effects.  
Evidently, even this expected enhancement is not enough, which increases the need for new observations to reveal the underlying physical processes.  Even if HAWC and LHAASO only set limits on the TeV gamma-ray flux, that would be important.

Our goal here is to provide a theoretical foundation to quantitatively assess the TeV detection prospects.  At low energies, cosmic rays are affected by magnetic modulation in the inner Solar System, as well as by magnetic fields in the solar atmosphere, all of which are complicated~\cite{Seckel:1991ffa}.  At high energies, where magnetic effects can be neglected, the calculations are relatively straightforward but have not been done before this paper.  The energy separating the two regimes is not known.  We estimate that neglecting magnetic effects is appropriate for TeV--PeV gamma rays and show that it leads to useful benchmarks for GeV--TeV gamma rays.  In future work, we will treat magnetic effects in detail.  For a broader context on our program of work on understanding the gamma-ray emission of the Sun --- aimed toward eventual new measurements of cosmic rays, among other goals --- see Ref.~\cite{Ng:2015gya}.

We now provide more information about gamma-ray observations and prospects.  Over the 0.1~GeV--TeV range, the Sun has been well observed.  Following the upper limits given by EGRET~\cite{JGRA:JGRA13592} and the first detection using EGRET archival data~\cite{Orlando:2008uk}, more detailed measurements were reported in Ref.~\cite{Abdo:2011xn} by the Fermi Collaboration, based on 1.5 years of data.  Over the range 0.1--10 GeV, they separately measured the disk and halo fluxes, finding spectra $\sim E^{-2}$, plus a hint of time variation in the disk flux.  In Ref.~\cite{Ng:2015gya}, where we used six years of Fermi data and a newer version of the data processing (Pass 7 vs Pass 6), we detected the disk flux up to 100~GeV, finding that its spectrum falls more steeply than $E^{-2}$.  We also made the first robust detection of time variation, showing that the disk flux decreased by a factor of 2.5 from solar minimum to maximum.  While the solar-halo gamma-ray flux is reasonably well understood, our results deepen the mysteries of the solar-disk gamma-ray flux.  New observations are needed, especially at higher energies, which will critically test emissions models.  However, this is difficult with Fermi due to the low gamma-ray flux.

In the TeV--PeV range, the only ground-based gamma-ray experiments that can observe the Sun are those that directly detect shower particles.  (For air-Cherenkov detectors, based on detecting optical photons, the Sun is too bright.)  The HAWC experiment began full operations in 2015, and is now reporting first results.  The LHAASO experiment, under construction, is expected to begin operations in 2020.  These experiments will greatly improve upon the energy range and flux sensitivity of their predecessors, e.g., Milagro~\cite{Atkins:1999gb}, ARGO-YBJ~ \cite{Aielli:2006cj}, and Tibet AS-gamma~\cite{Hibino:1988er}.  Those and other experiments have observed the ``Sun shadow," a deficit of shower particles caused by the solar disk blocking cosmic-ray hadrons~\cite{Amenomori:2013own, Enriquez:2015nva}, but none have detected an excess gamma-ray emission from the Sun.  The shadow is displaced by $\sim 1^\circ$ by from the Sun's position due to magnetic deflections of cosmic rays en route to Earth, but the gamma-ray excess will be centered on the Sun.  HAWC and LHAASO observations in the TeV range, combined with Fermi observations in the GeV range, will provide a long lever arm to test models of solar gamma-ray emission.

This paper makes steps toward a comprehensive understanding of solar gamma rays.  In Sec.~\ref{sec:B}, we discuss the effects of magnetic fields and justify why we can neglect them here.  The next three sections are ordered by the directionality of the signals.  In Sec.~\ref{sec:hadronic}, we detail our calculation of the hadronic gamma-ray emission from the limb of the solar disk.  {\it This calculation has not been done before.} We also estimate the flux of other secondary products (electrons, positrons, and neutrons), discussing if they are significant background for the gamma rays.  In Sec.~\ref{sec:leptonic}, we detail our calculation of the leptonic gamma-ray emission from the solar halo.  {\it We extend earlier calculations to higher energies and are the first to include allowed new contributions to the electron spectrum.}  In Sec.~\ref{sec_CREprobe}, we discuss the all-sky signal of directly detected cosmic-ray electrons.  {\it Our points about these prospects are new and exciting.}  In Sec.~\ref{sec:conclusions}, we present our conclusions and the outlook for further work.


\section{Interplanetary and Solar Magnetic Fields}
\label{sec:B}

The flux of cosmic rays near the Sun is altered by magnetic effects. Throughout the Solar System, there are magnetic disturbances sourced by the Sun and carried by the solar wind~\cite{Solanki:2010je,Priest2014a}. These form an interplanetary magnetic field (IMF) that repels galactic cosmic rays (``solar modulation")~\cite{jokipii1971,Potgieter:2013pdj}; the effects and their uncertainties increase at low energies and at small distances from the Sun.  In addition, near the Sun, within approximately $\sim 0.1$~AU, there are solar magnetic fields (SMF) that are quite strong, especially in the photosphere and corona~\cite{Wiegelmann2014a}.  Because the SMF are complex and not completely measured, their effects may be varied and are quite uncertain.

In this paper, we focus on gamma-ray signals in the energy regime where magnetic effects can be neglected.  When this is appropriate for the solar-disk signal, it will be even more so for the solar-halo signal, for which cosmic rays interact farther from the Sun.  We begin by discussing SMF effects on the solar-disk signal, as these turn out to be dominant over IMF effects.

SMF effects enhance gamma-ray production from the solar disk.  A likely physical mechanism was proposed in Ref.~\cite{Seckel:1991ffa}, although the authors' predictions still fall far below observations~\cite{Abdo:2011xn, Ng:2015gya}. The enhancement is due to the mirror effect of solar magnetic flux tubes on charged hadronic cosmic rays, which can reverse the directions of cosmic rays before they interact, thus producing outgoing gamma rays that are not absorbed by the Sun.  At high enough energies, this mirroring becomes ineffective, and the enhancement ends.  To estimate the critical energy $E_c$ for this transition, where magnetic-field effects on cosmic rays can be neglected, we compute the Larmor radius, $L$, using the typical SMF strength near the Sun, $B \sim 1$ G, and the solar radius, $R_\odot \simeq 7 \times10^{10}$ cm~\cite{Solanki:2010je, Priest2014a}, finding
\begin{equation}
E_c \sim 10^4 \, {{\rm GeV}} \left(\frac{L}{R_\odot}\right) \left(\frac{B}{1\, {\rm G}}\right).
\label{eq:Ec}
\end{equation}
A similar value is obtained for a single flux tube, for which the magnetic field strength can be $\sim 10^3$ times larger and the distance scale $\sim 10^3$ times smaller~\cite{Solanki:2010je, Priest2014a}. (Ref.~\cite{Seckel:1991ffa} estimated $E_c$ to be between $\simeq 3 \times10^2$~GeV and $\simeq2\times10^4$~GeV, so our choice is conservative.)  Because $E_\gamma \sim 0.1 E_p$ for typical hadronic interactions, SMF effects should therefore be negligible for gamma-ray energies above about 1 TeV.  However, SMF models are uncertain, and it is important to test them with new data.

IMF effects reduce gamma-ray fluxes.  Near Earth, IMF effects on the cosmic-ray spectrum are well described by the widely used force-field approximation~\cite{Gleeson:1968zza, Abdo:2011xn, Cholis:2015gna} and detailed simulations~\cite{Bobik:2011ig, Miyake:2006zp}, which are informed by extensive measurements.  For cosmic rays in the inner Solar System, both modeling and data are sparse.  A key clue is that the MESSENGER probe to Mercury found only $\lesssim 10\%$ modulation of the cosmic-ray spectrum above 0.125 GeV near solar distances around 0.4~AU~\cite{JGRA:JGRA51485, JGRA:JGRA52795}.  Using a force-field model with appropriate parameters to be consistent with these data (potentials $\lesssim 400 \, {\rm MV}$), we find that IMF effects can be neglected for cosmic rays with energies above 100~GeV (and thus gamma rays above 10~GeV), even near the solar surface.  However, IMF models are also uncertain, heightening the need for new data.

At energies where magnetic effects can be neglected, the solar-disk signal should thus be wholly due to the limb contribution.  This emission is caused by cosmic rays that graze the Sun, encountering a column density that is large enough for them to interact but small enough for their gamma rays to escape.  Because this signal can be calculated with minimal uncertainty, a gamma-ray measurement consistent with its flux prediction would confirm that magnetic effects are negligible.  In principle, this could also be checked by the angular distribution of the signal, where the Sun would appear as a bright ring with a dark center, although planned TeV--PeV experiments may not have adequate angular resolution~\cite{Abeysekara:2013tza, Abeysekara:2017mjj, Zhen:2014zpa, He:2016del}. Finally, tests could also be made by the time variation, as there should be none.

At lower energies, where magnetic effects are important, several distinctive signatures of the solar-disk signal should emerge.  The flux should be larger, as SMF effects that enhance the gamma-ray flux dominate over IMF effects that decrease it~\cite{Seckel:1991ffa}.  That is, our solar-disk prediction neglecting magnetic effects provides a theoretical {\it lower} bound on the disk flux, which is especially interesting at GeV--TeV energies.  The angular variation of the signal should tend toward illumination of the full disk.  And there should be time variations that reveal the nature of the dominant magnetic effects.  IMF effects decrease gamma-ray production near solar maximum, due to cosmic-ray modulation~\cite{Abdo:2011xn, Ng:2015gya}.  Perhaps surprisingly, SMF effects must act in the same sense, as the IMF effects are too small to explain the observed time variation~\cite{Ng:2015gya}.

For the solar-halo signal, IMF effects dominate over SMF effects~\cite{Abdo:2011xn}, except perhaps very near the Sun.  The comparison of disk and halo signals will thus help disentangle IMF and SMF effects.  It also means that neglecting magnetic effects provides a theoretical {\it upper} bound on the halo flux.


\section{Hadronic Gamma Rays}
\label{sec:hadronic}


\subsection{Calculational framework}
\label{sec:hadronic_framework}

In the direction of the solar disk, the dominant source of gamma rays is the interactions of hadronic cosmic rays with matter in the solar atmosphere~\cite{Seckel:1991ffa, Abdo:2011xn, Ng:2015gya}.  Of these interactions, the most important are inelastic proton-proton collisions that produce neutral pions, which promptly decay to gamma rays.  (In Sec.~\ref{sec:leptonic}, we calculate gamma-ray production by leptonic cosmic rays, including near the direction of the solar disk, although the interactions occur well away from the solar surface.)

Here we calculate the gamma-ray emission from the solar limb --- the small fraction of the Sun encountered by cosmic rays that just graze its surface on trajectories toward Earth.  We use the straight-line approximation, where gamma rays maintain the direction of their parent hadrons, appropriate because the particle energies are so high.  We ignore emission from the disk because we neglect magnetic effects that can reverse the directions of cosmic rays before they interact~\cite{Seckel:1991ffa} and because the contributions of back-scattered pions are tiny~\cite{Ambrosio:1997qh}. As the ingredients of the calculation are reasonably well known, the predicted limb emission is robust and, as noted, sets a theoretical lower bound on the solar-disk flux.

We calculate the total flux from the limb, integrating over its solid angle.  Here we assume that it cannot be resolved, as single-shower angular resolution of HAWC and LHAASO near 1 TeV is comparable to the $0.5^\circ$-diameter of the Sun~\cite{Abeysekara:2013tza, Abeysekara:2017mjj, Zhen:2014zpa, He:2016del}.  The solid angle of the limb is tiny, $\sim 10^{-3}$ of that of the solar disk.  If the limb could be resolved, it would appear as a thin, bright ring, with the intensity (flux per solid angle) enhanced by $\sim 10^3$ over the intensity averaged over the solar disk.  The angular resolutions of HAWC and LHAASO improve at higher energies, which may allow partial resolution of the limb, especially with stricter cuts to select events with the best angular resolution.  In the long term, hardware upgrades to improve this should be considered.

We begin in Sec.~\ref{sec:hadronic_simplified} by discussing gamma-ray production in a simplified case --- proton-proton production of neutral pions in the thin-target limit --- which can be handled semianalytically, following Ref.~\cite{Kelner:2006tc}.  Then, in Sec.~\ref{sec:hadronic_realistic}, we include the effects of multiple scattering and absorption, cascade processes, and nuclear composition through a simulation using {\tt GEANT4}~\cite{Agostinelli:2002hh, Geant4}.  In ths simplified case, the flux is
\begin{multline}
   \frac{dF}{dE_\gamma}(E_\gamma) =
   \int d\Omega \int ds\, n_p(\vec{s}) \\
   \int dE_p\, \frac{dI}{dE_p}(E_p) \, \sigma_{\rm inel}(E_p) \, \frac{dN}{dE_\gamma}(E_p,E_\gamma),\label{eq:pp_thin}
\end{multline}
where $n_p$ is the number density of target protons at the line-of-sight coordinate $\vec{s}$, $dI/dE_p$ is the cosmic-ray proton intensity, $\sigma_{\rm inel}$ is the inelastic proton-proton scattering cross section, and $dN/d{E_\gamma}$ is the spectrum of gamma rays per interaction.  The length of the chord through the solar atmosphere is $\Delta s \sim (8 R_\odot h_0)^{1/2} \sim 2.6 \times 10^4$~km, where the 8 comes from geometry, $R_\odot \simeq 7 \times 10^{5}$~km is the radius of the Sun, and $h_0 \simeq 120$ km is the scale height of the solar matter density in the photosphere.  In the realistic case, the most important interactions occur at proton optical depths $\tau \sim 2$, so this simplified case is not adequate for our full results, although it does introduce the framework well.


\subsection{Calculation for the simplified case}
\label{sec:hadronic_simplified}

\begin{figure}[t]
\includegraphics[width=\columnwidth]{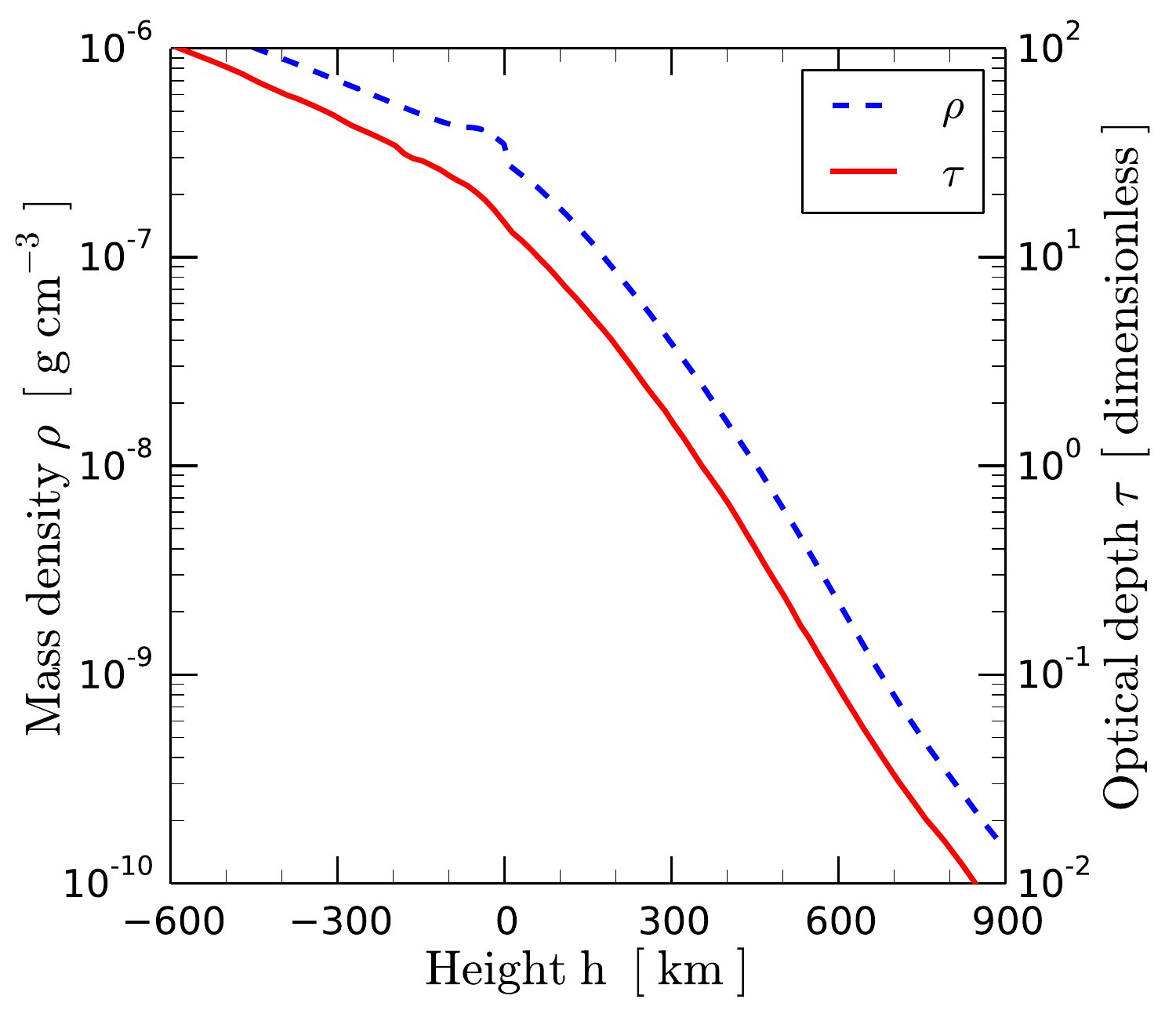}
\caption{Solar mass density as function of height above the photosphere (left axis, blue dashed), as well as the same for optical depth for inelastic proton-proton collisions (right axis, red solid).
}
\label{fig:density}
\end{figure}

Figure~\ref{fig:density} shows the solar mass density $\rho$ from Refs.~\cite{baker1966tables, 1973ApJ...184..605V}. Above the photosphere, the density declines exponentially, following the Boltzmann distribution of gravitational potential energy in the nearly isothermal atmosphere.  Figure~\ref{fig:density} also shows the proton optical depth $\tau$ as a function of height above the photosphere.  The cross section for inelastic proton-proton collisions changes only modestly with energy and is $\simeq 30 - 70$~mb for proton energies $1 - 10^7$ GeV~\cite{Agashe:2014kda}.  In the optically thin limit, gamma-ray production is dominated by the decay of neutral pions, which, at the low densities considered here, always decay in flight before interacting.  Kelner, Aharonian, and Bugayov~\cite{Kelner:2006tc} have extensively studied the yields of secondaries in proton-proton collisions, in which their results are based on a fit to data and to particle-interaction simulations.  The yield of gamma rays has a broad energy spectrum, but the most important gamma ray typically has $E_\gamma \sim 0.1 E_p$.  The shape of $\tau(h)$ closely follows that of $\rho(h)$, due to the exponential dependence, with the conversion factor $\simeq \sigma_{\rm inel} s / m_p \simeq 5 \times 10^7 \, {\rm cm^3 \, g^{-1}}$.

The cosmic-ray flux can be taken to be that at Earth, as we neglect magnetic effects.  (Technically, the flux at Earth includes some modulation effects, but these are negligible at such high energies.)  Up to 1 TeV, we use the precisely measured proton spectrum from the Alpha Magnetic Spectrometer (AMS-02)~\cite{Aguilar:2015ooa}.  At higher energies, it is sufficient to extrapolate this using $dI/dE \simeq 1 \, (E/{\rm GeV})^{-2.7}$ cm$^{-2}$ s$^{-1}$ sr$^{-1}$ GeV$^{-1}$~\cite{Agashe:2014kda}.

Figure~\ref{fig:hadronic} shows the resulting gamma-ray spectrum for the case where we integrated over
$0 < \tau < 0.3$, up to roughly the largest value for which an optically thin calculation is appropriate (the probability for a proton to interact twice is then $\lesssim 10\%$, so the gamma-ray spectrum scales linearly with $\tau$).  We checked the results of our semianalytic calculation by a Monte Carlo simulation with the particle-interaction code {\tt GEANT4}~\cite{Agostinelli:2002hh, Geant4}, for which the results matched to within $\lesssim 10\%$.  This shows that effects beyond those in Ref.~\cite{Kelner:2006tc}, e.g., particle cascades in the medium, are unimportant.

Lastly, compared to the Sun, the gamma-ray flux from the limb of the Earth's atmosphere has been measured by Fermi up to nearly 1 TeV and compared to simulations, finding good agreement with predictions, which demonstrates the robustness of theoretical calculations~\cite{Abdo:2009gt, Ackermann:2014ula}. In principle, in the thin-target limit, the limb flux from the Sun could simply be expressed in terms of the limb flux from Earth, nullifying several potential uncertainties, such as the energy spectrum, composition, and cross section.

\subsection{Calculation for the realistic case}
\label{sec:hadronic_realistic}

To include proton-proton interactions in the optically thick case, we use {\tt GEANT4}~\cite{Agostinelli:2002hh, Geant4}.  This allows protons to interact several times, and takes into account their particle and energy losses from all processes.  It also includes gamma-ray production by cascade processes, such as bremsstrahlung by electrons.  The number density of target photons is $\sim10^4$ times smaller than that of solar matter so energy losses and gamma-ray production by inverse-Compton processes can be neglected~\cite{Rott:2012qb}.
The density is low enough that charged pions below 1 PeV will typically decay in flight before interacting.  Neutrons and muons may escape, and the neutrons may survive to Earth without decay.

\begin{figure}[t]
    \includegraphics[width=\columnwidth]{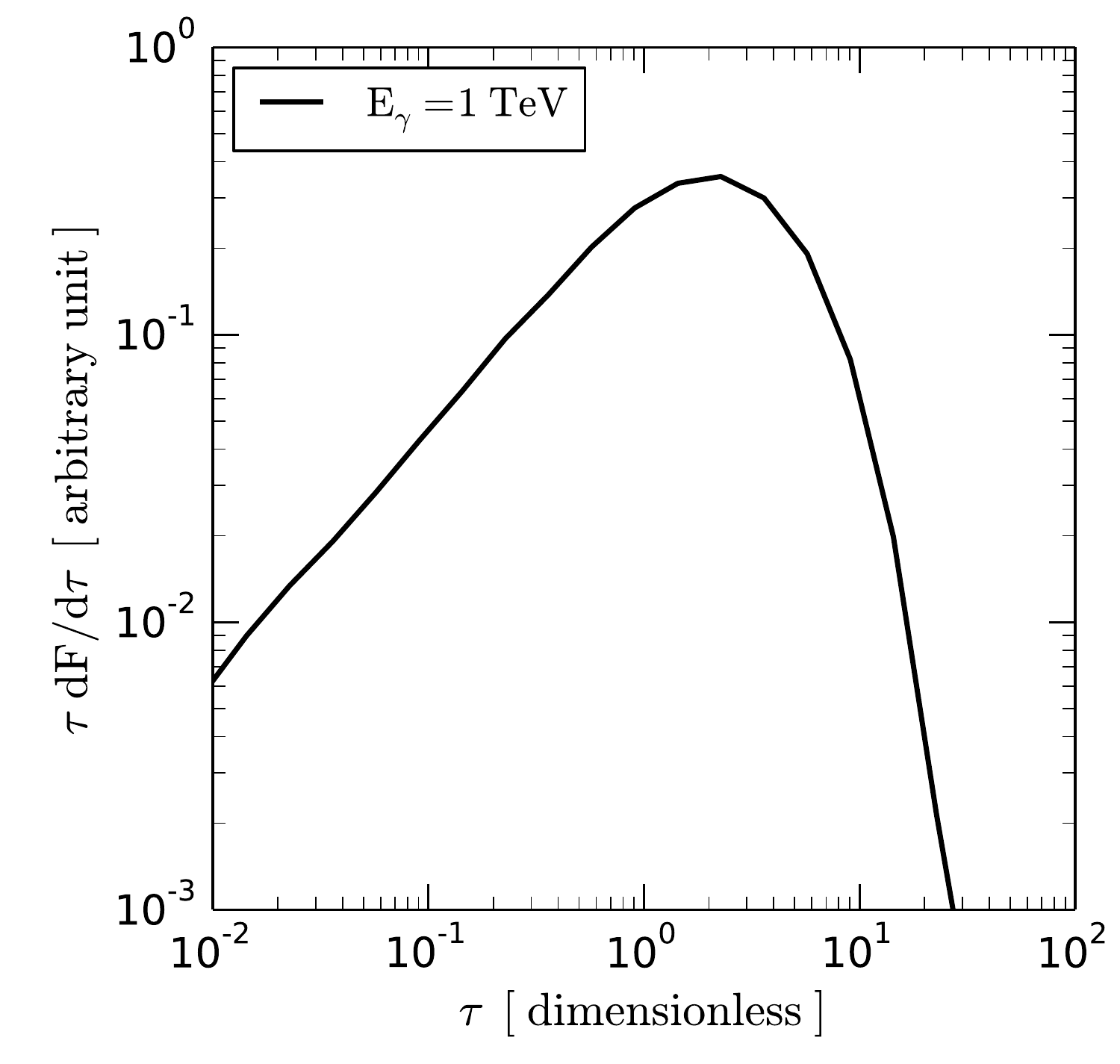}
\caption{Normalized relative contributions of different $\tau$ values to the predicted gamma-ray flux, based on our {\tt GEANT4} simulation.  We show the example of $E_\gamma=1$~TeV; other energies give similar results.
}
\label{fig:dFdtau}
\end{figure}

Figure~\ref{fig:dFdtau} shows the range of $\tau$ values that contribute most to gamma-ray production, based on our {\tt GEANT4} simulation.  The y axis is weighted to properly compare different logarithmic ranges of $\tau$.  The peak is near $\tau \sim 2$, where $\sim 90\%$ of cosmic-ray protons will interact at least once.  To the left of the peak, the linear decline is due to reduced optical depth.  To the right, the exponential decline is due to proton cooling and especially gamma-ray absorption, which happen to have similar interaction lengths (for pion production and electron-positron pair production, respectively).  In combination, about 90\% of the total flux arises in the range $\tau \sim 0.1-10$.

\begin{figure}[t]
\includegraphics[width=\columnwidth]{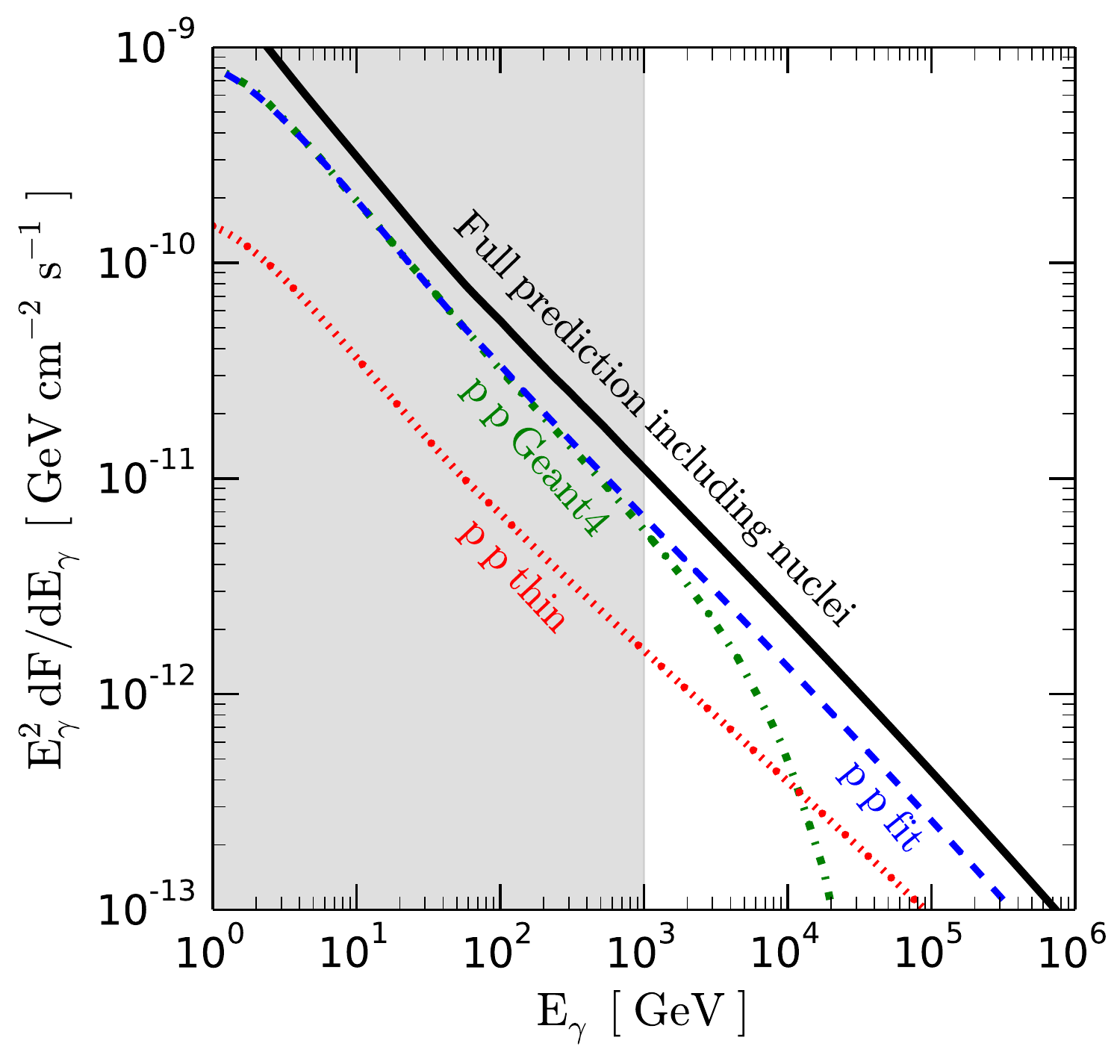}
\caption{Solar-limb gamma-ray spectrum produced by hadronic cosmic rays.  Red dotted line: semianalytic result for proton-proton interactions with $0 < \tau < 0.3$.  Green dash-dotted line: {\tt GEANT4} results for the full range of $\tau$; the gradual cutoff is because it cannot simulate proton interactions above 100 TeV.  Blue dashed line: our empirical fit to the {\tt GEANT4} results, extrapolated to higher energies.  Black solid line: our full prediction, including a correction factor for nuclei.  The light grey shading approximately indicates the energies at which magnetic effects, neglected here, should be included.
}
\label{fig:hadronic}
\end{figure}

Using this range of $\tau$ values, we use Fig.~\ref{fig:density} to determine the corresponding range of heights above the photosphere and corresponding mass densities, finding $h \sim 60 - 600$~km and $\rho \sim 10^{-7} - 10^{-9}$~g~cm$^{-3}$.  This leads to important insights about the physical conditions in which interactions occur.  In this range, the solar properties are reasonably well known and are stable in time.  The conditions for the production of solar atmospheric gamma rays are quite different from those for Earth atmospheric neutrinos~\cite{Gaisser:2002jj}; for the latter, $\rho \sim 10^{-4}$ g cm$^{-3}$ at an altitude of 10 km, and the distance scales are short but the proton optical depth is high ($\tau \sim 20$).  Lastly, this information will be useful for assessing interactions in the presence of magnetic effects, which we will consider in future work.

With {\tt GEANT4}, we can simulate proton interactions only up to a laboratory energy of 100 TeV, which leads to a gradual cutoff of the gamma-ray spectrum near 10 TeV.  To extend our results to higher energies, we develop an empirical fit to the {\tt GEANT4} results at lower energies.  We modify our semianalytic approach, Eq.~(\ref{eq:pp_thin}), by including a correction factor, $e^{-\alpha\tau(E_p,\vec{s})}$, that only becomes important in the optically thick regime.  For the free parameter $\alpha$, we find that 0.65 gives a good match to the {\tt GEANT4} results.  This is shown in Fig.~\ref{fig:hadronic}.

Finally, we consider the effect of nuclei in the cosmic rays and in the solar atmosphere.  Besides protons, the only important constituent is helium, which has a $\simeq 10\%$ relative number abundance in both the beam and target~\cite{baker1966tables, 1973ApJ...184..605V, Agashe:2014kda}.  We use the cosmic-ray helium data from Ref.~\cite{Aguilar:2015ctt} up to 1~TeV.  Above that, we use a power law and extrapolate up to $\sim10$ PeV with spectral index 2.7, which roughly describes the data compilation in Ref.~\cite{Agashe:2014kda}.
Following Ref.~\cite{Kachelriess:2014mga}, we calculate the gamma-ray flux enhancement factor due to cosmic-ray helium. We find that the gamma-ray flux is increased by an overall factor $\simeq 1.8$, with a small energy dependence due to a slightly different spectral shape between the proton and helium.
We also consider the case in which the helium spectrum may be harder than 2.7 at high energies~\cite{Ahn:2010gv,Yoon:2011aa,Atkin:2017vhi}.  If we use a spectral index of 2.58~\cite{Yoon:2011aa} for extrapolation, our result changes by less than 20\% near 10\,TeV.  Thus, we can safely ignore the spectral hardening.

Figure~\ref{fig:hadronic} shows our full prediction for the gamma-ray spectrum from the solar limb.  The gamma-ray spectrum closely follows the cosmic-ray proton spectrum~\cite{Kelner:2006tc}.  This is because the pions and gamma rays typically carry fixed fractions of the parent proton energy, the cross sections and multiplicities for (high-energy) pion production and gamma-ray absorption have only mild energy dependence, and the pions decay before interacting.  (For the same reasons, Earth atmospheric neutrinos at sub-TeV energies also follow the cosmic-ray spectrum~\cite{Gaisser:2002jj}.)  In Fig.~\ref{fig:hadronic}, the gamma-ray flux has normalization $\simeq 2 \times 10^{-9}$ sr times the proton intensity (flux per solid angle).  This factor can be roughly reproduced using $\Delta \Omega \times \tau \times (0.1)^{1.7}$, where $\Delta \Omega \sim 10^{-7}$ sr is the relevant solid angle of the limb, $\tau \sim 1$ is a typical value, and the last factor comes from assuming that each proton produces $\sim 1$ gamma-ray at $E_\gamma \sim 0.1 E_p$.

The hadronic-interaction processes discussed here also produce neutrinos, electrons (including positrons), and neutrons~\cite{Seckel:1991ffa}.  The neutrino flux~\cite{Moskalenko:1991hm, Moskalenko:1993ke, Ingelman:1996mj, Fogli:2006jk} is an important background for dark matter searches with neutrino telescopes~\cite{Tanaka:2011uf, Aartsen:2012kia, Rott:2011fh}, and constitutes a sensitivity floor~\cite{Arguelles:2017eao, Ng:2017aur, Edsjo:2017kjk}.  The other species could be useful messengers to study cosmic-ray interactions with the Sun, using detectors such as Fermi~\cite{FermiLAT:2011ab}, AMS-02~\cite{Aguilar:2013qda}, CALET~\cite{Torii:2011zza}, and DAMPE~\cite{ChangJin:550}.  A dedicated study of their detectability, is beyond the scope of this paper, and will be considered elsewhere.  Here we briefly comment on their relevance to gamma-ray observations.

Electrons can be effectively separated from gamma rays in space-borne detectors.  However, this separation is difficult for ground-based experiments, as both particles induce electromagnetic showers in the atmosphere.  In principle, the inclusion of electrons enhances the detectability of the Sun for ground-based experiments. The flux of the electrons can be estimated similarly to that of gamma rays, described above, also by first ignoring magnetic-field effects.  The electron flux is found to $\sim 2$ times lower than that of the gamma rays, due to receiving a smaller fraction of the pion energy.  Further, the detection of these secondary electrons with ground-based experiments is more complicated than gamma rays,  as the effects of solar, interplanetary, and Earth magnetic fields all need to be taken into account, demonstrated by cosmic-ray shadow studies~\cite{Amenomori:2013own, Enriquez:2015nva}.  The deflections and diffusion they cause will reduce the electron flux per solid angle.  Therefore, for the current study, we neglect the addition of the electron flux to the total electromagnetic signal observable by ground-based experiments.

Neutrons, the most important secondary hadrons, travel without being affected by the magnetic fields.  The Sun is therefore a point source of neutrons, and could in principle be detectable by ground-based experiments.  Compared to gamma-ray production in pionic processes, secondary neutrons carry a smaller fraction of the primary energy.  However, spallation of helium is efficient at producing secondary neutrons.  Combining these two factors, the limb neutron flux is comparable to that of the gamma rays~(also the disk flux~\cite{Seckel:1991ffa}).  In practice, it is difficult for these neutrons to be confused with gamma rays by ground-based experiments, due to the excellent hadron rejection factor, $\sim 10^{-3}$.  The detection in the hadron channel is also likely to be difficult due to the much higher background, compared to that of gamma rays and electrons.  A more careful treatment of hadrons, in particular at lower energies, is the subject of a separate paper (Zhou {\it et al.}, in prep.).


\section{Leptonic Gamma Rays}
\label{sec:leptonic}

In directions away from the solar disk, there is a solar halo of gamma-ray emission, of which the dominant source is the interactions of cosmic-ray electrons ($e^++e^-$) with solar photons~\cite{Moskalenko:2006ta, Orlando:2006zs, Orlando:2008uk, Abdo:2011xn}.  Of these interactions, the most important is inverse-Compton scattering.  There is also a contribution in the direction of the solar disk.  We estimate that other interactions with solar photons are irrelevant; these include Bethe-Heitler~\cite{heitler1954quantum, Berezinsky:2002nc} and photo-pion interactions of protons~\cite{Andersen:2011dz} and deexcitation interactions of nuclei following photodisintegration~\cite{Karakula:1994nv, Anchordoqui:2006pd, Anchordoqui:2006pe, Murase:2010va}.

Here we calculate this leptonic gamma-ray emission, mostly following prior work~\cite{Moskalenko2000a, Moskalenko:2006ta, Orlando:2006zs, Orlando:2008uk}. For the first time, we calculate results up to 1 PeV and show that uncertainties in the electron spectrum at very high energies allow larger signals than in the nominal case (a broken power-law spectrum for cosmic-ray electrons).  As above, we neglect magnetic effects and assume straight-line propagation for the parent-daughter kinematics.  Although the solar halo flux is present in all directions, its intensity (flux per solid angle) is greatest near the Sun, falling approximately as $\theta^{-1}$~\cite{Moskalenko:2006ta, Orlando:2006zs, Orlando:2008uk} , where $\theta$ is the angle away from the center of the Sun.  The flux within a given angle thus grows as $\theta$, but the backgrounds --- especially significant for ground-based detectors --- grow as $\theta^2$.  Therefore, the solar-halo signal is most interesting at relatively small angles.
We calculate the leptonic signal within $1.5^\circ$ degrees of the solar center; this value matches what we used for our Fermi analysis~\cite{Ng:2015gya} and will allow HAWC and LHAASO to treat it as a near-point source.

In the optically thin regime, the gamma-ray flux from the inverse-Compton interactions of cosmic-ray electrons is
\begin{multline}
   \frac{dF}{d{E_\gamma}} = \int d\Omega \int ds
   \int d{E_{ph}} \, \frac {d n_{ph}} {d{E_{ph}}} ({E_{ph}}, \vec{s}) \\
   \int dE_e \, \frac {d\sigma} {d{E_\gamma}} (E_e, {E_{ph}}, {E_\gamma})
   \frac {dI} {dE_e} (E_e, \vec{s})
\label{eq:IC}
\end{multline}
where $dn_{ph}/d{E_{ph}}(\vec{s})$ is the number-density spectrum of target photons at the line-of-sight coordinate $\vec{s}$, $dI/dE_e$ is the cosmic-ray intensity, and $d\sigma/d{E_\gamma}$ is the electron-photon differential cross section including Klein-Nishina effects.

\begin{figure}[t]
\includegraphics[width=\columnwidth]{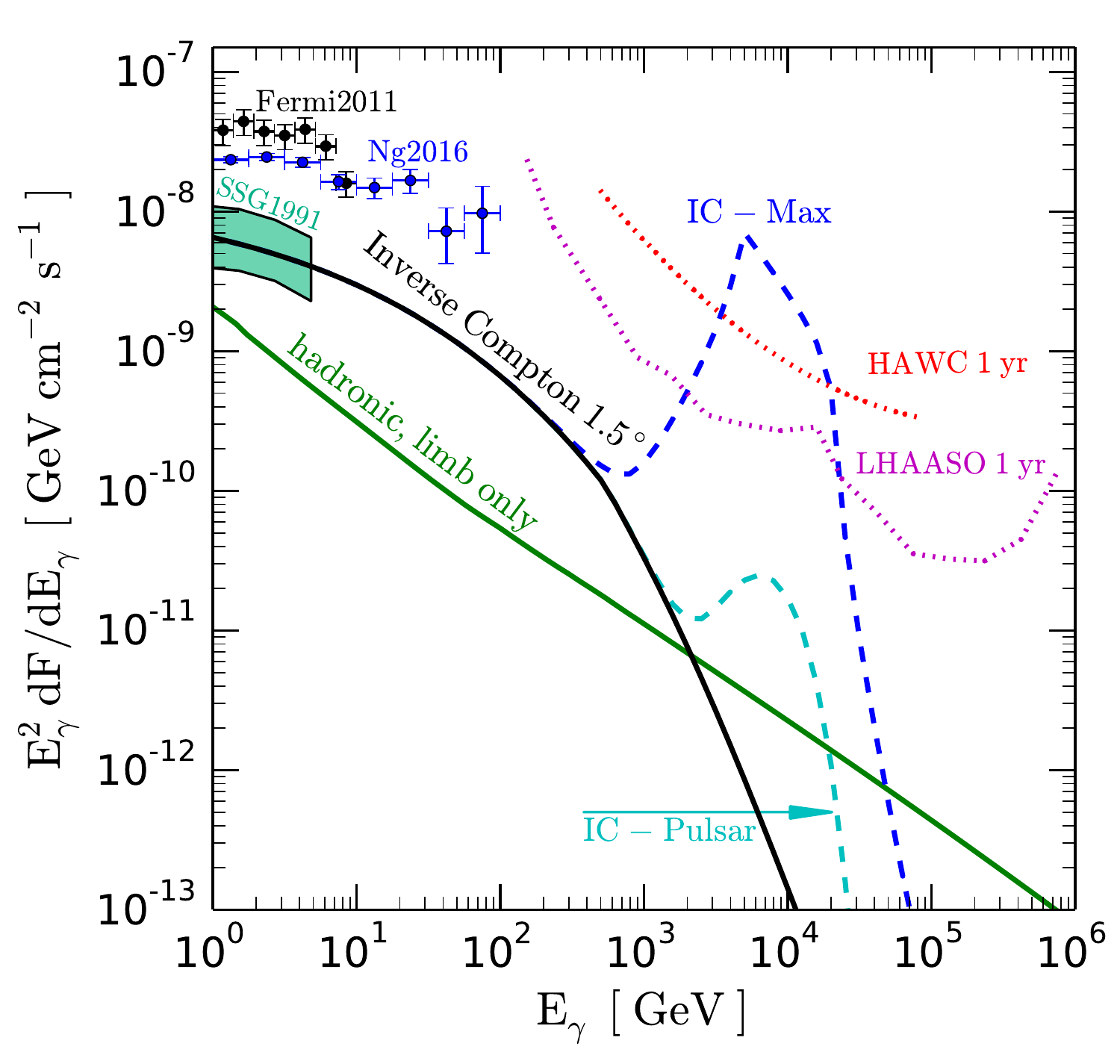}
\caption{Gamma-ray spectrum of the Sun.  Points: disk observations with Fermi~\cite{Abdo:2011xn, Ng:2015gya}, where the flux difference is due to time variation.  Green band: the predicted disk flux~\cite{Seckel:1991ffa}.  Dotted lines: the estimated differential point-source sensitivity of HAWC~\cite{Abeysekara:2017mjj} (scaled to one year) and LHAASO~\cite{Zhen:2014zpa, He:2016del}.  Our new prediction of the solar-disk signal due to cosmic-ray hadrons (from the limb) is shown by the green solid line.  Our new prediction of the solar-halo signal due to inverse-Compton scattering of cosmic-ray electrons is shown by the black solid line for the nominal case and by the dashed lines for enhanced cases from Fig.~\ref{fig:espec}.
}
\label{fig:detectability}
\end{figure}

The column density of the solar photon field is $\sim n_{ph} R_\odot^2 / D \theta$ for small angles $\theta$~\cite{Orlando:2008uk}, where $n_{ph}$ is the number density of photons at the solar surface and $D = 1$ AU.  For electron energies below about 0.25 TeV, the inverse-Compton cross section is in the Thompson regime, where the total cross section is constant with energy.  At higher energies, it is in the Klein-Nishina regime, where the total cross section falls with increasing energy.  An electron passing close to the Sun has an optical depth of $\sim 10^{-2}$ (in the Thompson regime; less at higher energies), so the optically thin assumption of Eq.~(\ref{eq:IC}) is appropriate.  To calculate the gamma-ray spectrum, we use the {\tt StellarICs} code~\cite{Orlando:2013pza, Orlando:2013nga}, slightly modified to include a parametrization of the electron spectrum at the highest energies.  The solar photons are taken to have a blackbody spectrum with temperature 5780~K and corresponding typical energy of $\sim 1$~eV.  The photon density falls as distance squared far from the Sun but less quickly near its surface, where it varies as with radial distance $r$ as $[1 - (1 - R_\odot^2 / r^2)]$~\cite{Moskalenko:2006ta, Orlando:2008uk}.  The cosmic-ray electron flux has been precisely measured by AMS-02 up to almost 1 TeV~\cite{Aguilar:2014mma}, and measured moderately well by H.E.S.S.~\cite{Aharonian:2008aa, Aharonian:2009ah} and VERITAS~\cite{Staszak:2015kza} up to 5 TeV.  We use a broken power-law fit to these data.  As discussed in detail in Sec.~\ref{sec_CREprobe}, the electron spectrum at very high energies might be much larger than expected from this nominal case, in which the flux above 5 TeV is assumed to fall off quickly.  Our calculation is the first to show how allowed contributions to the electron spectrum above 5 TeV would enhance the solar-halo gamma-ray signal.


\section{Cosmic-Ray Electrons}
\label{sec_CREprobe}

\begin{figure}[t]
\includegraphics[width=\columnwidth]{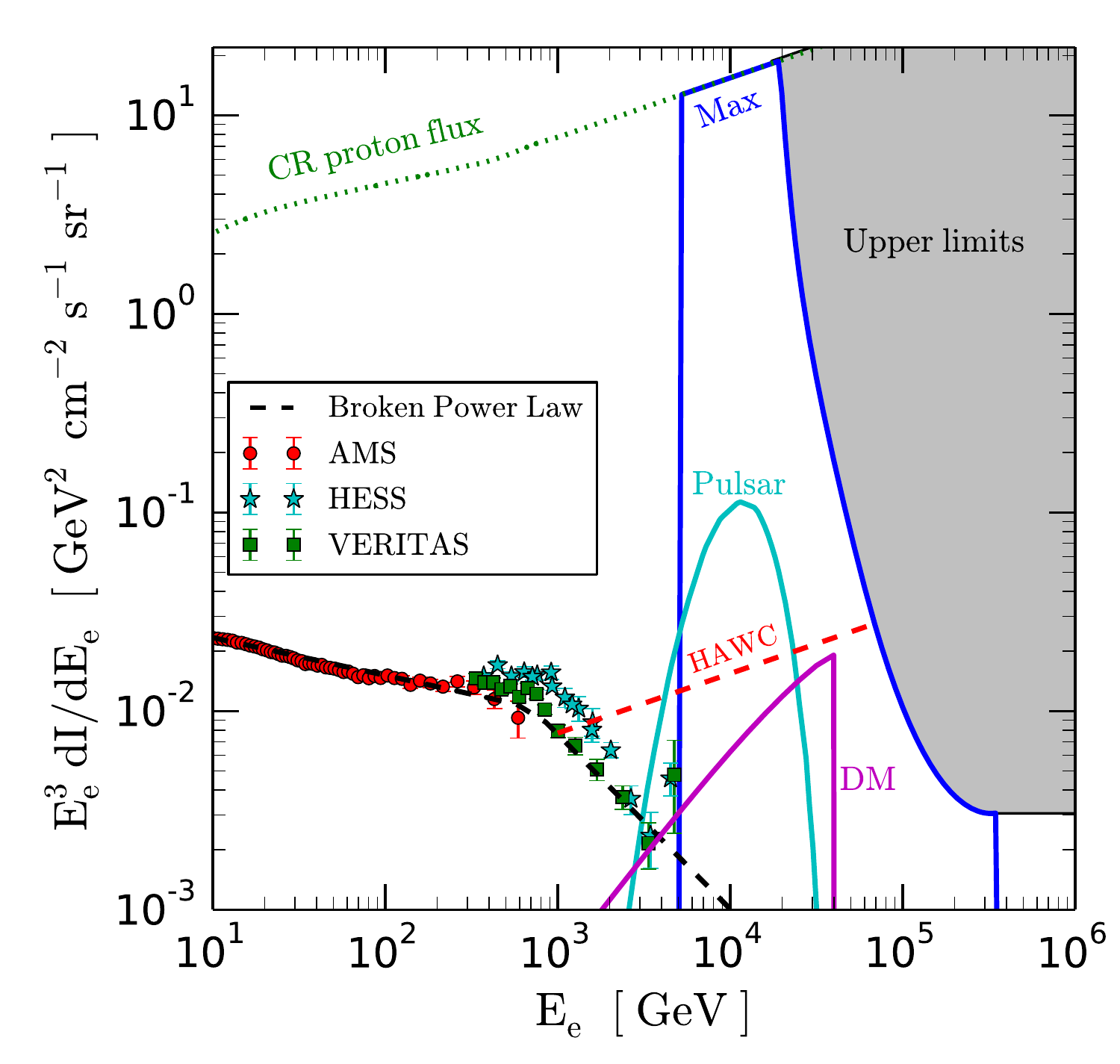}
\caption{Diffuse flux (weighted with $E_e^3$) of cosmic-ray electrons.  Below about 5 TeV, there are measurements (points, as labeled~\cite{Aguilar:2014mma, Aharonian:2009ah, Aharonian:2008aa, Staszak:2015kza}).  Above about 70 TeV, there are limits (gray region, which combines many experiments~\cite{Kistler:2009wm, Ahlers:2013xia}).  In between, the spectrum could be as large as the blue solid line, allowing enhanced contributions (pulsar or dark matter; details are in the text). HAWC should be able to immediately improve sensitivity down to $\sim 10^{-3}$ (hadronic rejection) of the proton spectrum (red dashed line).}
\label{fig:espec}
\end{figure}

Figure~\ref{fig:detectability} shows our results for the leptonic gamma-ray emission in the nominal case plus some enhanced cases.  (Below 10 GeV, where there are measurements from Fermi~\cite{Abdo:2011xn}, not shown here, our prediction is consistent.)  In the Thomson regime, the gamma-ray spectrum is less steep than the electron spectrum due to the nature of the differential cross section.  In the Klein-Nishina regime, the gamma-ray spectrum steepens sharply due to the suppression of the total cross section (in addition to the steepening electron spectrum).  The nominal predictions are not detectable with HAWC and LHAASO.  In fact, only the most extreme enhanced scenarios --- with the cosmic-ray electron flux as large as the proton flux --- are (lines labeled ``Max'' in Figs.~\ref{fig:detectability} and~\ref{fig:espec}).  If no solar-halo signals are detected, as is likely, that will make it easier to isolate hadronic gamma-ray flux in the direction of the solar disk.  Section~\ref{sec_CREprobe} introduces a better way to probe cosmic-ray electrons.

Figure~\ref{fig:detectability} also recaps our result for the hadronic gamma-ray emission from the solar limb.  This is well below the leptonic gamma-ray emission from the solar halo near the disk (below about 1 TeV), as well as the sensitivity of HAWC and LHAASO.  However, this prediction leads to several important points.  The gamma rays observed from the solar disk must be hadronic, with their flux enhanced by magnetic effects, and the ratio of the data to our limb prediction provides a first direct measure of the strength of that enhancement.  The hadronic gamma-ray spectrum must eventually bend toward and join with our limb prediction.  Until the energy at which that occurs, there is positive evidence for interesting processes (magnetic effects) beyond the limb emission.  It may be that the leptonic gamma-ray emission is never dominant in the data, despite its apparent dominance in Fig.~\ref{fig:detectability}.

Here we show that HAWC and LHAASO can directly measure the cosmic-ray electron ($e^++e^-$) spectrum, which is of great interest~\cite{Chang:2008aa, Aharonian:2008aa, Adriani:2008zr, Profumo:2008ms, Aharonian:2009ah, Kistler:2009wm, Hinton:2011ad, FermiLAT:2011ab, Torii:2011zza, Adriani:2011xv, Aguilar:2013qda, Ahlers:2013xia, ChangJin:550, Staszak:2015kza}.  Compared to the method of Sec.~\ref{sec:leptonic}, this is simpler and more powerful.  The flux is expected to be isotropic.  If a nearby pulsar or dark matter halo contributes significantly, the resulting anisotropy would enhance the detection prospects, but we neglect this possibility.  Because cosmic-ray electrons lose energy quickly, by synchrotron and inverse-Compton processes, the highest-energy electrons must come from quite nearby, e.g., a few hundred pc at 10 TeV.

Figure~\ref{fig:espec} summarizes present knowledge of the cosmic-ray electron spectrum.  Below 5 TeV, there are measurements from various detectors, including AMS-02~\cite{Aguilar:2014mma}, H.E.S.S.~\cite{Aharonian:2008aa, Aharonian:2009ah}, and VERITAS~\cite{Staszak:2015kza}.  Above 70 TeV, there are strong limits from ground-based arrays (summarized in Refs.~\cite{Kistler:2009wm, Ahlers:2013xia}).  Importantly, at 5--70 TeV, there have been no experimental probes, as emphasized in Ref.~\cite{Kistler:2009wm}.  At those energies, the only limit, which is quite weak, comes from requiring that the electron flux not exceed the all-particle flux.  New sensitivity is needed to probe the electron spectrum in this energy range, where new components could appear.  Intriguingly, there are hints of a new component starting to emerge at 5 TeV, seen by both the southern-sky H.E.S.S.~\cite{Aharonian:2008aa} and the northern-sky VERITAS~\cite{Staszak:2015kza}.

HAWC and LHAASO detect electrons and gamma rays with comparable efficiency~\cite{BenZvi:2015kga}.  However, the flux sensitivity for electrons is worse because, like the background protons, they are isotropic.  The sensitivity depends on just the hadronic rejection factor.  (Gamma rays are not a background, except in the direction of point sources;  the diffuse flux of TeV electrons, even in the nominal case, exceeds that of gamma rays, even in the direction of the Milky Way plane~\cite{Prodanovic:2006bq}.)  We assume a hadronic rejection factor of $\sim 10^3$, which should be reachable (Segev BenZvi, private communication).  Performance close to this has been demonstrated by some analyses with a partially complete HAWC detector~\cite{Solares:2015jza, Pretz:2015zja}. More importantly, HAWC has already shown preliminary limits that approach our estimated sensitivity~\cite{Pretz:2015wma}.

Figure~\ref{fig:espec} shows the estimated HAWC sensitivity to the electron flux (LHAASO's will likely be similar), along with possible enhancements to the 5--70 TeV electron spectrum.  HAWC and LHAASO can reach higher energies than air-Cherenkov detectors because of their huge advantages in field of view and uptime.

Probing the 5--70 TeV cosmic-ray electron spectrum for the first time will allow interesting tests of pulsars, dark matter, and possible surprises.  For pulsars, we use predictions from Refs.~\cite{Profumo:2008ms, Hinton:2011ad}, which may explain the positron excess~\cite{Adriani:2008zr, Aguilar:2013qda}.  (Even larger fluxes can be found in Ref.~\cite{Fang:2016wid}.)  For dark matter, we use the {\tt PPPC4DMID} code~\cite{Cirelli:2010xx, Buch:2015iya} to calculate the electron spectra from dark matter decay, in this case with a mass of 100 TeV and a lifetime of $2 \times10^{26}$~s, which is comparable to current constraints~\cite{Murase:2012xs, Esmaili:2013gha, Feldstein:2013kka}.

While simple, our results are important.  Although the gap in coverage of the cosmic-ray electron spectrum was known~\cite{Kistler:2009wm}, as was the possibility of using HAWC to detect electrons~\cite{BenZvi:2015kga}, this paper is the first to combine those points and quantify the prospects.  In the near future, there will be good sensitivity to high-energy cosmic-ray electrons from the CALET~\cite{Torii:2011zza}, DAMPE~\cite{ChangJin:550} and CTA~\cite{Consortium:2010bc} experiments.  Even so, they may only reach $\lesssim20$~TeV. With more than a year of data already collected, HAWC has a unique opportunity now, and we encourage swift action to complete an analysis.


\section{Conclusions and Outlook}
\label{sec:conclusions}

The Sun's high-energy gamma-ray emission --- seemingly due to irradiation by cosmic rays --- is not well understood.  Above 10 GeV, the Sun is one of the $\sim 20$ brightest sources detected by Fermi, and its disk emission is nearly an order of magnitude brighter~\cite{Abdo:2011xn, Ng:2015gya} than predicted~\cite{Seckel:1991ffa}.  In the TeV range, there have been no theoretical or observational studies.

Now there is a convergence of two opportunities: the recognition that the high-energy Sun can reveal important physics and the unprecedented sensitivity of the already running HAWC experiment.  These opportunities will be enhanced by ongoing theoretical work and the sensitivity gain due to the coming LHAASO experiment.

This paper has three main results.

{\it The first calculation of the gamma-ray emission due to hadronic cosmic rays interacting with the solar limb.}
At high enough energies ($\gtrsim 1$ TeV), magnetic effects can be neglected, and the complete emission from the solar disk should be from only the thin ring of the limb.  This flux can be robustly calculated.  Further, it serves as an important theoretical lower bound on the solar-disk emission at all energies.  The enhancement of the disk flux by magnetic fields can be deduced by the ratio of the observed flux to this prediction.  In the GeV range, this is a factor $\sim 10$.  As illustrated in Fig.~\ref{fig:intro}, HAWC and LHAASO will provide new sensitivity to solar gamma rays in the TeV range, and can test if this enhancement continues, plus if there are new contributions, e.g., due to dark matter. (Limits from ARGO-YBJ~\cite{Aielli:2006cj} are already in preparation~\cite{zheli}, and is about one order of magnitude weaker than HAWC sensitivity at TeV energies.) Finally, the limb flux would be significantly more detectable if the solar disk could be resolved, due to lower backgrounds per solid angle.  Although we have conservatively neglected this possibility, it seems attainable.

{\it New results on the gamma-ray emission due to cosmic-ray electrons interacting with solar photons.}
This emission forms a gamma-ray halo around the Sun, and the intensity peaks near the disk.  For the first time, we calculate the TeV--PeV gamma-ray flux, including the possibility of new components in the 5--70 TeV electron spectrum.  HAWC and LHAASO can at least set constraints at these energies, where there are no measurements.

{\it A new perspective on allowed enhancements to the cosmic-ray electron spectrum and direct tests of such.}
Lastly, we show that direct observations of electromagnetic showers by HAWC and LHAASO can provide unprecedented sensitivity to the 5--70 TeV cosmic-ray electron spectrum.  This search, based on nondirectional signals, will be a powerful probe of the high-energy electron spectrum, testing some realistic models.

This paper is part of a larger program of work to develop the Sun as a new high-energy laboratory (see Ref.~\cite{Ng:2015gya} for further discussion).  With a good theoretical understanding of magnetic effects, the Sun could be used as a passive detector for cosmic rays in the inner Solar System, allowing measurements that are differential in particle type and energy, a capability unmatched by any existing or planned detector.  Currently, the major roadblock to this goal is taking into account the complicated magnetic field effects, but this problem is tractable in principle, and progress is being made (Zhou {\it et al.}, in preparation).  The Sun is already a calibration source for direction, and could become one for flux.  Interestingly, unlike any other astrophysical source, the Sun's hadronic and leptonic emission can be clearly separated using angular information alone.  Finally, a thorough understanding of cosmic-ray interactions with the Sun is crucial for testing dark matter and neutrino physics~\cite{Leane:2017vag, Arina:2017sng}.


\begin{acknowledgments}
We thank Andrea Albert, Mauricio Bustamante, Rebecca Leane, Shirley Li, Shoko Miyake, Carsten Rott, Qingwen Tang, and especially Segev BenZvi, Igor Moskalenko, Elena Orlando, and Andrew Strong for helpful discussions.
BZ was supported by Ohio State University's Fowler and University Fellowships.  KCYN was supported by NASA Grant No. NNX13AP49G, Ohio State's Presidential Fellowship, and NSF Grant No. PHY-1404311.  JFB was supported by NSF Grant No. PHY-1404311.  AHGP was supported by NASA Grant No. NNX13AP49G.
\end{acknowledgments}

\bibliography{references}

\begin{thebibliography}{98}%
\makeatletter
\providecommand \@ifxundefined [1]{%
 \@ifx{#1\undefined}
}%
\providecommand \@ifnum [1]{%
 \ifnum #1\expandafter \@firstoftwo
 \else \expandafter \@secondoftwo
 \fi
}%
\providecommand \@ifx [1]{%
 \ifx #1\expandafter \@firstoftwo
 \else \expandafter \@secondoftwo
 \fi
}%
\providecommand \natexlab [1]{#1}%
\providecommand \enquote  [1]{``#1''}%
\providecommand \bibnamefont  [1]{#1}%
\providecommand \bibfnamefont [1]{#1}%
\providecommand \citenamefont [1]{#1}%
\providecommand \href@noop [0]{\@secondoftwo}%
\providecommand \href [0]{\begingroup \@sanitize@url \@href}%
\providecommand \@href[1]{\@@startlink{#1}\@@href}%
\providecommand \@@href[1]{\endgroup#1\@@endlink}%
\providecommand \@sanitize@url [0]{\catcode `\\12\catcode `\$12\catcode
  `\&12\catcode `\#12\catcode `\^12\catcode `\_12\catcode `\%12\relax}%
\providecommand \@@startlink[1]{}%
\providecommand \@@endlink[0]{}%
\providecommand \url  [0]{\begingroup\@sanitize@url \@url }%
\providecommand \@url [1]{\endgroup\@href {#1}{\urlprefix }}%
\providecommand \urlprefix  [0]{URL }%
\providecommand \Eprint [0]{\href }%
\providecommand \doibase [0]{http://dx.doi.org/}%
\providecommand \selectlanguage [0]{\@gobble}%
\providecommand \bibinfo  [0]{\@secondoftwo}%
\providecommand \bibfield  [0]{\@secondoftwo}%
\providecommand \translation [1]{[#1]}%
\providecommand \BibitemOpen [0]{}%
\providecommand \bibitemStop [0]{}%
\providecommand \bibitemNoStop [0]{.\EOS\space}%
\providecommand \EOS [0]{\spacefactor3000\relax}%
\providecommand \BibitemShut  [1]{\csname bibitem#1\endcsname}%
\let\auto@bib@innerbib\@empty
\bibitem [{\citenamefont {Seckel}\ \emph {et~al.}(1991)\citenamefont {Seckel},
  \citenamefont {Stanev},\ and\ \citenamefont {Gaisser}}]{Seckel:1991ffa}%
  \BibitemOpen
  \bibfield  {author} {\bibinfo {author} {\bibfnamefont {D.}~\bibnamefont
  {Seckel}}, \bibinfo {author} {\bibfnamefont {Todor}\ \bibnamefont {Stanev}},
  \ and\ \bibinfo {author} {\bibfnamefont {T.~K.}\ \bibnamefont {Gaisser}},\
  }\bibfield  {title} {\enquote {\bibinfo {title} {{Signatures of cosmic-ray
  interactions on the solar surface}},}\ }\href {\doibase 10.1086/170753}
  {\bibfield  {journal} {\bibinfo  {journal} {Astrophys. J.}\ }\textbf
  {\bibinfo {volume} {382}},\ \bibinfo {pages} {652--666} (\bibinfo {year}
  {1991})}\BibitemShut {NoStop}%
\bibitem [{\citenamefont {Moskalenko}\ \emph {et~al.}(2006)\citenamefont
  {Moskalenko}, \citenamefont {Porter},\ and\ \citenamefont
  {Digel}}]{Moskalenko:2006ta}%
  \BibitemOpen
  \bibfield  {author} {\bibinfo {author} {\bibfnamefont {Igor~V.}\ \bibnamefont
  {Moskalenko}}, \bibinfo {author} {\bibfnamefont {Troy~A.}\ \bibnamefont
  {Porter}}, \ and\ \bibinfo {author} {\bibfnamefont {Seth~W.}\ \bibnamefont
  {Digel}},\ }\bibfield  {title} {\enquote {\bibinfo {title} {{Inverse Compton
  scattering on solar photons, heliospheric modulation, and neutrino
  astrophysics}},}\ }\href {\doibase 10.1086/520882, 10.1086/509916} {\bibfield
   {journal} {\bibinfo  {journal} {Astrophys. J.}\ }\textbf {\bibinfo {volume}
  {652}},\ \bibinfo {pages} {L65--L68} (\bibinfo {year} {2006})},\ \bibinfo
  {note} {[Erratum: Astrophys. J. 664, L143 (2007)]},\ \Eprint
  {http://arxiv.org/abs/astro-ph/0607521} {arXiv:astro-ph/0607521 [astro-ph]}
  \BibitemShut {NoStop}%
\bibitem [{\citenamefont {Orlando}\ and\ \citenamefont
  {Strong}(2007)}]{Orlando:2006zs}%
  \BibitemOpen
  \bibfield  {author} {\bibinfo {author} {\bibfnamefont {Elena}\ \bibnamefont
  {Orlando}}\ and\ \bibinfo {author} {\bibfnamefont {Andrew}\ \bibnamefont
  {Strong}},\ }\bibfield  {title} {\enquote {\bibinfo {title} {{Gamma-rays from
  halos around stars and the Sun}},}\ }\bibfield  {booktitle} {\emph {\bibinfo
  {booktitle} {{The Multi-Messenger Approach to High-Energy Gamma-Ray Sources:
  3rd Workshop on the Nature of Unidentified High-Energy Sources, Barcelona,
  Spain, 4-7 Jul, 2006}}},\ }\href {\doibase 10.1007/s10509-007-9457-0}
  {\bibfield  {journal} {\bibinfo  {journal} {Astrophys. Space Sci.}\ }\textbf
  {\bibinfo {volume} {309}},\ \bibinfo {pages} {359--363} (\bibinfo {year}
  {2007})},\ \Eprint {http://arxiv.org/abs/astro-ph/0607563}
  {arXiv:astro-ph/0607563 [astro-ph]} \BibitemShut {NoStop}%
\bibitem [{\citenamefont {Orlando}\ and\ \citenamefont
  {Strong}(2008)}]{Orlando:2008uk}%
  \BibitemOpen
  \bibfield  {author} {\bibinfo {author} {\bibfnamefont {Elena}\ \bibnamefont
  {Orlando}}\ and\ \bibinfo {author} {\bibfnamefont {Andrew~W.}\ \bibnamefont
  {Strong}},\ }\bibfield  {title} {\enquote {\bibinfo {title} {{Gamma-ray
  emission from the solar halo and disk: a study with EGRET data}},}\ }\href
  {\doibase 10.1051/0004-6361:20078817} {\bibfield  {journal} {\bibinfo
  {journal} {Astron. Astrophys.}\ }\textbf {\bibinfo {volume} {480}},\ \bibinfo
  {pages} {847} (\bibinfo {year} {2008})},\ \Eprint
  {http://arxiv.org/abs/0801.2178} {arXiv:0801.2178 [astro-ph]} \BibitemShut
  {NoStop}%
\bibitem [{\citenamefont {Abdo}\ \emph {et~al.}(2011)\citenamefont {Abdo} \emph
  {et~al.}}]{Abdo:2011xn}%
  \BibitemOpen
  \bibfield  {author} {\bibinfo {author} {\bibfnamefont {A.~A.}\ \bibnamefont
  {Abdo}} \emph {et~al.} (\bibinfo {collaboration} {Fermi-LAT Collaboration}),\
  }\bibfield  {title} {\enquote {\bibinfo {title} {{Fermi-LAT Observations of
  Two Gamma-Ray Emission Components from the Quiescent Sun}},}\ }\href
  {\doibase 10.1088/0004-637X/734/2/116} {\bibfield  {journal} {\bibinfo
  {journal} {Astrophys. J.}\ }\textbf {\bibinfo {volume} {734}},\ \bibinfo
  {pages} {116} (\bibinfo {year} {2011})},\ \Eprint
  {http://arxiv.org/abs/1104.2093} {arXiv:1104.2093 [astro-ph.HE]} \BibitemShut
  {NoStop}%
\bibitem [{\citenamefont {Ng}\ \emph {et~al.}(2016)\citenamefont {Ng},
  \citenamefont {Beacom}, \citenamefont {Peter},\ and\ \citenamefont
  {Rott}}]{Ng:2015gya}%
  \BibitemOpen
  \bibfield  {author} {\bibinfo {author} {\bibfnamefont {Kenny C.~Y.}\
  \bibnamefont {Ng}}, \bibinfo {author} {\bibfnamefont {John~F.}\ \bibnamefont
  {Beacom}}, \bibinfo {author} {\bibfnamefont {Annika H.~G.}\ \bibnamefont
  {Peter}}, \ and\ \bibinfo {author} {\bibfnamefont {Carsten}\ \bibnamefont
  {Rott}},\ }\bibfield  {title} {\enquote {\bibinfo {title} {{First Observation
  of Time Variation in the Solar-Disk Gamma-Ray Flux with Fermi}},}\ }\href
  {\doibase 10.1103/PhysRevD.94.023004} {\bibfield  {journal} {\bibinfo
  {journal} {Phys. Rev.}\ }\textbf {\bibinfo {volume} {D94}},\ \bibinfo {pages}
  {023004} (\bibinfo {year} {2016})},\ \Eprint
  {http://arxiv.org/abs/1508.06276} {arXiv:1508.06276 [astro-ph.HE]}
  \BibitemShut {NoStop}%
\bibitem [{\citenamefont {Schneid}\ \emph {et~al.}(1996)\citenamefont {Schneid}
  \emph {et~al.}}]{Schneid1996a}%
  \BibitemOpen
  \bibfield  {author} {\bibinfo {author} {\bibfnamefont {E.~J.}\ \bibnamefont
  {Schneid}} \emph {et~al.},\ }\bibfield  {title} {\enquote {\bibinfo {title}
  {{EGRET observations of X-class solar flares.}}}\ }\href@noop {} {\bibfield
  {journal} {\bibinfo  {journal} {Astronomy and Astrophysics Supplement
  Series}\ }\textbf {\bibinfo {volume} {120}},\ \bibinfo {pages} {299--302}
  (\bibinfo {year} {1996})}\BibitemShut {NoStop}%
\bibitem [{\citenamefont {Ajello}\ \emph {et~al.}(2014)\citenamefont {Ajello}
  \emph {et~al.}}]{Fermi-LAT:2013cla}%
  \BibitemOpen
  \bibfield  {author} {\bibinfo {author} {\bibfnamefont {M.}~\bibnamefont
  {Ajello}} \emph {et~al.} (\bibinfo {collaboration} {Fermi-LAT
  Collaboration}),\ }\bibfield  {title} {\enquote {\bibinfo {title} {{Impulsive
  and Long Duration High-Energy Gamma-ray Emission From the Very Bright 2012
  March 7 Solar Flares}},}\ }\href {\doibase 10.1088/0004-637X/789/1/20}
  {\bibfield  {journal} {\bibinfo  {journal} {Astrophys. J.}\ }\textbf
  {\bibinfo {volume} {789}},\ \bibinfo {pages} {20} (\bibinfo {year} {2014})},\
  \Eprint {http://arxiv.org/abs/1304.5559} {arXiv:1304.5559 [astro-ph.HE]}
  \BibitemShut {NoStop}%
\bibitem [{\citenamefont {Ackermann}\ \emph
  {et~al.}(2014{\natexlab{a}})\citenamefont {Ackermann} \emph
  {et~al.}}]{Ackermann:2014rma}%
  \BibitemOpen
  \bibfield  {author} {\bibinfo {author} {\bibfnamefont {M.}~\bibnamefont
  {Ackermann}} \emph {et~al.} (\bibinfo {collaboration} {Fermi-LAT
  Collaboration}),\ }\bibfield  {title} {\enquote {\bibinfo {title}
  {{High-energy Gamma-Ray Emission from Solar Flares: Summary of Fermi Large
  Area Telescope Detections and Analysis of Two M-class Flares}},}\ }\href
  {\doibase 10.1088/0004-637X/787/1/15} {\bibfield  {journal} {\bibinfo
  {journal} {Astrophys. J.}\ }\textbf {\bibinfo {volume} {787}},\ \bibinfo
  {pages} {15} (\bibinfo {year} {2014}{\natexlab{a}})},\ \Eprint
  {http://arxiv.org/abs/1304.3749} {arXiv:1304.3749 [astro-ph.HE]} \BibitemShut
  {NoStop}%
\bibitem [{\citenamefont {Pesce-Rollins}\ \emph {et~al.}(2015)\citenamefont
  {Pesce-Rollins}, \citenamefont {Omodei}, \citenamefont {Petrosian},
  \citenamefont {Liu}, \citenamefont {da~Costa}, \citenamefont {Allafort},\
  and\ \citenamefont {Chen}}]{Pesce-Rollins:2015hpa}%
  \BibitemOpen
  \bibfield  {author} {\bibinfo {author} {\bibfnamefont {Melissa}\ \bibnamefont
  {Pesce-Rollins}}, \bibinfo {author} {\bibfnamefont {Nicola}\ \bibnamefont
  {Omodei}}, \bibinfo {author} {\bibfnamefont {Vahe'}\ \bibnamefont
  {Petrosian}}, \bibinfo {author} {\bibfnamefont {Wei}\ \bibnamefont {Liu}},
  \bibinfo {author} {\bibfnamefont {Fatima~Rubio}\ \bibnamefont {da~Costa}},
  \bibinfo {author} {\bibfnamefont {Alice}\ \bibnamefont {Allafort}}, \ and\
  \bibinfo {author} {\bibfnamefont {Qingrong}\ \bibnamefont {Chen}},\
  }\bibfield  {title} {\enquote {\bibinfo {title} {{First detection of $>$100
  MeV gamma rays associated with a behind-the-limb solar flare}},}\ }\href
  {\doibase 10.1088/2041-8205/805/2/L15} {\bibfield  {journal} {\bibinfo
  {journal} {Astrophys. J.}\ }\textbf {\bibinfo {volume} {805}},\ \bibinfo
  {pages} {L15} (\bibinfo {year} {2015})},\ \Eprint
  {http://arxiv.org/abs/1505.03480} {arXiv:1505.03480 [astro-ph.SR]}
  \BibitemShut {NoStop}%
\bibitem [{\citenamefont {Abeysekara}\ \emph {et~al.}(2017)\citenamefont
  {Abeysekara} \emph {et~al.}}]{Abeysekara:2017mjj}%
  \BibitemOpen
  \bibfield  {author} {\bibinfo {author} {\bibfnamefont {A.~U.}\ \bibnamefont
  {Abeysekara}} \emph {et~al.},\ }\bibfield  {title} {\enquote {\bibinfo
  {title} {{Observation of the Crab Nebula with the HAWC Gamma-Ray
  Observatory}},}\ }\href@noop {} {\  (\bibinfo {year} {2017})},\ \Eprint
  {http://arxiv.org/abs/1701.01778} {arXiv:1701.01778 [astro-ph.HE]}
  \BibitemShut {NoStop}%
\bibitem [{\citenamefont {Cao}(2014)}]{Zhen:2014zpa}%
  \BibitemOpen
  \bibfield  {author} {\bibinfo {author} {\bibfnamefont {Zhen}\ \bibnamefont
  {Cao}},\ }\bibfield  {title} {\enquote {\bibinfo {title} {{LHAASO: Science
  and Status}},}\ }\bibfield  {booktitle} {\emph {\bibinfo {booktitle}
  {{Proceedings, Vulcano Workshop 2014: Frontier Objects in Astrophysics and
  Particle Physics: Vulcano, Italy, May 18-24, 2014}}},\ }\href@noop {}
  {\bibfield  {journal} {\bibinfo  {journal} {Frascati Phys. Ser.}\ }\textbf
  {\bibinfo {volume} {58}},\ \bibinfo {pages} {331} (\bibinfo {year}
  {2014})}\BibitemShut {NoStop}%
\bibitem [{\citenamefont {He}(2016)}]{He:2016del}%
  \BibitemOpen
  \bibfield  {author} {\bibinfo {author} {\bibfnamefont {Huihai}\ \bibnamefont
  {He}} (\bibinfo {collaboration} {LHAASO Collaboration}),\ }\bibfield  {title}
  {\enquote {\bibinfo {title} {{Design highlights and status of the LHAASO
  project}},}\ }\bibfield  {booktitle} {\emph {\bibinfo {booktitle}
  {{Proceedings, 34th International Cosmic Ray Conference (ICRC 2015): The
  Hague, The Netherlands, July 30-August 6, 2015}}},\ }\href@noop {} {\bibfield
   {journal} {\bibinfo  {journal} {PoS}\ }\textbf {\bibinfo {volume}
  {ICRC2015}},\ \bibinfo {pages} {1010} (\bibinfo {year} {2016})}\BibitemShut
  {NoStop}%
\bibitem [{\citenamefont {Rodgers}\ \emph {et~al.}(2015)\citenamefont
  {Rodgers}, \citenamefont {Lawrence}, \citenamefont {Feldman},\ and\
  \citenamefont {Peplowski}}]{JGRA:JGRA51485}%
  \BibitemOpen
  \bibfield  {author} {\bibinfo {author} {\bibfnamefont {Douglas~J.}\
  \bibnamefont {Rodgers}}, \bibinfo {author} {\bibfnamefont {David~J.}\
  \bibnamefont {Lawrence}}, \bibinfo {author} {\bibfnamefont {William~C.}\
  \bibnamefont {Feldman}}, \ and\ \bibinfo {author} {\bibfnamefont
  {Patrick~N.}\ \bibnamefont {Peplowski}},\ }\bibfield  {title} {\enquote
  {\bibinfo {title} {{Neutrons and energetic charged particles in the inner
  heliosphere: Measurements of the MESSENGER Neutron Spectrometer from 0.3 to
  0.85~AU}},}\ }\href {\doibase 10.1002/2014JA020263} {\bibfield  {journal}
  {\bibinfo  {journal} {Journal of Geophysical Research: Space Physics}\
  }\textbf {\bibinfo {volume} {120}},\ \bibinfo {pages} {841--854} (\bibinfo
  {year} {2015})}\BibitemShut {NoStop}%
\bibitem [{\citenamefont {Lawrence}\ \emph {et~al.}(2016)\citenamefont
  {Lawrence}, \citenamefont {Peplowski}, \citenamefont {Feldman}, \citenamefont
  {Schwadron},\ and\ \citenamefont {Spence}}]{JGRA:JGRA52795}%
  \BibitemOpen
  \bibfield  {author} {\bibinfo {author} {\bibfnamefont {David~J.}\
  \bibnamefont {Lawrence}}, \bibinfo {author} {\bibfnamefont {Patrick~N.}\
  \bibnamefont {Peplowski}}, \bibinfo {author} {\bibfnamefont {William~C.}\
  \bibnamefont {Feldman}}, \bibinfo {author} {\bibfnamefont {Nathan~A.}\
  \bibnamefont {Schwadron}}, \ and\ \bibinfo {author} {\bibfnamefont
  {Harlan~E.}\ \bibnamefont {Spence}},\ }\bibfield  {title} {\enquote {\bibinfo
  {title} {{Galactic cosmic ray variations in the inner heliosphere from solar
  distances less than 0.5~AU: Measurements from the MESSENGER Neutron
  Spectrometer}},}\ }\href {\doibase 10.1002/2016JA022962} {\bibfield
  {journal} {\bibinfo  {journal} {Journal of Geophysical Research: Space
  Physics}\ }\textbf {\bibinfo {volume} {121}},\ \bibinfo {pages} {7398--7406}
  (\bibinfo {year} {2016})}\BibitemShut {NoStop}%
\bibitem [{\citenamefont {{Jokipii}}(1971)}]{jokipii1971}%
  \BibitemOpen
  \bibfield  {author} {\bibinfo {author} {\bibfnamefont {J.~R.}\ \bibnamefont
  {{Jokipii}}},\ }\bibfield  {title} {\enquote {\bibinfo {title} {{Propagation
  of cosmic rays in the solar wind.}}}\ }\href {\doibase
  10.1029/RG009i001p00027} {\bibfield  {journal} {\bibinfo  {journal} {Reviews
  of Geophysics and Space Physics}\ }\textbf {\bibinfo {volume} {9}},\ \bibinfo
  {pages} {27--87} (\bibinfo {year} {1971})}\BibitemShut {NoStop}%
\bibitem [{\citenamefont {Potgieter}(2013)}]{Potgieter:2013pdj}%
  \BibitemOpen
  \bibfield  {author} {\bibinfo {author} {\bibfnamefont {Marius}\ \bibnamefont
  {Potgieter}},\ }\bibfield  {title} {\enquote {\bibinfo {title} {{Solar
  Modulation of Cosmic Rays}},}\ }\href {\doibase 10.12942/lrsp-2013-3}
  {\bibfield  {journal} {\bibinfo  {journal} {Living Rev. Solar Phys.}\
  }\textbf {\bibinfo {volume} {10}},\ \bibinfo {pages} {3} (\bibinfo {year}
  {2013})},\ \Eprint {http://arxiv.org/abs/1306.4421} {arXiv:1306.4421
  [physics.space-ph]} \BibitemShut {NoStop}%
\bibitem [{\citenamefont {Thompson}\ \emph {et~al.}(1997)\citenamefont
  {Thompson}, \citenamefont {Bertsch}, \citenamefont {Morris},\ and\
  \citenamefont {Mukherjee}}]{JGRA:JGRA13592}%
  \BibitemOpen
  \bibfield  {author} {\bibinfo {author} {\bibfnamefont {D.~J.}\ \bibnamefont
  {Thompson}}, \bibinfo {author} {\bibfnamefont {D.~L.}\ \bibnamefont
  {Bertsch}}, \bibinfo {author} {\bibfnamefont {D.~J.}\ \bibnamefont {Morris}},
  \ and\ \bibinfo {author} {\bibfnamefont {R.}~\bibnamefont {Mukherjee}},\
  }\bibfield  {title} {\enquote {\bibinfo {title} {Energetic gamma ray
  experiment telescope high-energy gamma ray observations of the moon and quiet
  sun},}\ }\href {\doibase 10.1029/97JA01045} {\bibfield  {journal} {\bibinfo
  {journal} {Journal of Geophysical Research: Space Physics}\ }\textbf
  {\bibinfo {volume} {102}},\ \bibinfo {pages} {14735--14740} (\bibinfo {year}
  {1997})}\BibitemShut {NoStop}%
\bibitem [{\citenamefont {Atkins}\ \emph {et~al.}(2000)\citenamefont {Atkins}
  \emph {et~al.}}]{Atkins:1999gb}%
  \BibitemOpen
  \bibfield  {author} {\bibinfo {author} {\bibfnamefont {Robert~W.}\
  \bibnamefont {Atkins}} \emph {et~al.} (\bibinfo {collaboration} {Milagro
  Collaboration}),\ }\bibfield  {title} {\enquote {\bibinfo {title}
  {{Milagrito: A TeV air shower array}},}\ }\href {\doibase
  10.1016/S0168-9002(00)00146-7} {\bibfield  {journal} {\bibinfo  {journal}
  {Nucl. Instrum. Meth.}\ }\textbf {\bibinfo {volume} {A449}},\ \bibinfo
  {pages} {478--499} (\bibinfo {year} {2000})},\ \Eprint
  {http://arxiv.org/abs/astro-ph/9912456} {arXiv:astro-ph/9912456 [astro-ph]}
  \BibitemShut {NoStop}%
\bibitem [{\citenamefont {Aielli}\ \emph {et~al.}(2006)\citenamefont {Aielli}
  \emph {et~al.}}]{Aielli:2006cj}%
  \BibitemOpen
  \bibfield  {author} {\bibinfo {author} {\bibfnamefont {G.}~\bibnamefont
  {Aielli}} \emph {et~al.} (\bibinfo {collaboration} {Argo-YBJ}),\ }\bibfield
  {title} {\enquote {\bibinfo {title} {{Layout and performance of RPCs used in
  the Argo-YBJ experiment}},}\ }\href {\doibase 10.1016/j.nima.2006.02.136}
  {\bibfield  {journal} {\bibinfo  {journal} {Nucl. Instrum. Meth.}\ }\textbf
  {\bibinfo {volume} {A562}},\ \bibinfo {pages} {92--96} (\bibinfo {year}
  {2006})}\BibitemShut {NoStop}%
\bibitem [{\citenamefont {Hibino}\ \emph {et~al.}(1989)\citenamefont {Hibino}
  \emph {et~al.}}]{Hibino:1988er}%
  \BibitemOpen
  \bibfield  {author} {\bibinfo {author} {\bibfnamefont {K.}~\bibnamefont
  {Hibino}} \emph {et~al.} (\bibinfo {collaboration} {TIBET ASgamma
  Collaboration}),\ }\bibfield  {title} {\enquote {\bibinfo {title} {{A New
  project to search for high-energy gamma-ray point sources in Tibet.}}}\
  }\bibfield  {booktitle} {\emph {\bibinfo {booktitle} {{Proceedings of the
  Workshop on High Resolution Gamma-Ray Cosmology November 2-5, 1988, Los
  Angeles}}},\ }\href {\doibase 10.1016/0920-5632(89)90057-1} {\bibfield
  {journal} {\bibinfo  {journal} {Nucl. Phys. Proc. Suppl.}\ }\textbf {\bibinfo
  {volume} {10B}},\ \bibinfo {pages} {219--227} (\bibinfo {year}
  {1989})}\BibitemShut {NoStop}%
\bibitem [{\citenamefont {Amenomori}\ \emph {et~al.}(2013)\citenamefont
  {Amenomori} \emph {et~al.}}]{Amenomori:2013own}%
  \BibitemOpen
  \bibfield  {author} {\bibinfo {author} {\bibfnamefont {M.}~\bibnamefont
  {Amenomori}} \emph {et~al.} (\bibinfo {collaboration} {TIBET ASgamma
  Collaboration}),\ }\bibfield  {title} {\enquote {\bibinfo {title} {{Probe of
  the Solar Magnetic Field Using the "Cosmic-Ray Shadow" of the Sun}},}\
  }\href {\doibase 10.1103/PhysRevLett.111.011101} {\bibfield  {journal}
  {\bibinfo  {journal} {Phys. Rev. Lett.}\ }\textbf {\bibinfo {volume} {111}},\
  \bibinfo {pages} {011101} (\bibinfo {year} {2013})},\ \Eprint
  {http://arxiv.org/abs/1306.3009} {arXiv:1306.3009 [astro-ph.SR]} \BibitemShut
  {NoStop}%
\bibitem [{\citenamefont {Enriquez-Rivera}\ and\ \citenamefont
  {Lara}(2016)}]{Enriquez:2015nva}%
  \BibitemOpen
  \bibfield  {author} {\bibinfo {author} {\bibfnamefont {Olivia}\ \bibnamefont
  {Enriquez-Rivera}}\ and\ \bibinfo {author} {\bibfnamefont {Alejandro}\
  \bibnamefont {Lara}} (\bibinfo {collaboration} {HAWC Collaboration}),\
  }\bibfield  {title} {\enquote {\bibinfo {title} {{The Galactic cosmic-ray Sun
  shadow observed by HAWC}},}\ }\bibfield  {booktitle} {\emph {\bibinfo
  {booktitle} {{Proceedings, 34th International Cosmic Ray Conference (ICRC
  2015): The Hague, The Netherlands, July 30-August 6, 2015}}},\ }\href@noop {}
  {\bibfield  {journal} {\bibinfo  {journal} {PoS}\ }\textbf {\bibinfo {volume}
  {ICRC2015}},\ \bibinfo {pages} {099} (\bibinfo {year} {2016})},\ \Eprint
  {http://arxiv.org/abs/1508.07351} {arXiv:1508.07351 [astro-ph.SR]}
  \BibitemShut {NoStop}%
\bibitem [{\citenamefont {Solanki}\ \emph {et~al.}(2006)\citenamefont
  {Solanki}, \citenamefont {Inhester},\ and\ \citenamefont
  {Schussler}}]{Solanki:2010je}%
  \BibitemOpen
  \bibfield  {author} {\bibinfo {author} {\bibfnamefont {Sami~K.}\ \bibnamefont
  {Solanki}}, \bibinfo {author} {\bibfnamefont {Bernd}\ \bibnamefont
  {Inhester}}, \ and\ \bibinfo {author} {\bibfnamefont {Manfred}\ \bibnamefont
  {Schussler}},\ }\bibfield  {title} {\enquote {\bibinfo {title} {{The solar
  magnetic field}},}\ }\href {\doibase 10.1088/0034-4885/69/3/R02} {\bibfield
  {journal} {\bibinfo  {journal} {Rept. Prog. Phys.}\ }\textbf {\bibinfo
  {volume} {69}},\ \bibinfo {pages} {563--668} (\bibinfo {year} {2006})},\
  \Eprint {http://arxiv.org/abs/1008.0771} {arXiv:1008.0771 [astro-ph.SR]}
  \BibitemShut {NoStop}%
\bibitem [{\citenamefont {Priest}(2014)}]{Priest2014a}%
  \BibitemOpen
  \bibfield  {author} {\bibinfo {author} {\bibfnamefont {Eric}\ \bibnamefont
  {Priest}},\ }\href@noop {} {\emph {\bibinfo {title} {Magnetohydrodynamics of
  the Sun}}}\ (\bibinfo  {publisher} {Cambridge University Press},\ \bibinfo
  {year} {2014})\BibitemShut {NoStop}%
\bibitem [{\citenamefont {Wiegelmann}\ \emph {et~al.}(2014)\citenamefont
  {Wiegelmann}, \citenamefont {Thalmann},\ and\ \citenamefont
  {Solanki}}]{Wiegelmann2014a}%
  \BibitemOpen
  \bibfield  {author} {\bibinfo {author} {\bibfnamefont {Thomas}\ \bibnamefont
  {Wiegelmann}}, \bibinfo {author} {\bibfnamefont {Julia~K.}\ \bibnamefont
  {Thalmann}}, \ and\ \bibinfo {author} {\bibfnamefont {Sami~K.}\ \bibnamefont
  {Solanki}},\ }\bibfield  {title} {\enquote {\bibinfo {title} {{The magnetic
  field in the solar atmosphere}},}\ }\href {\doibase
  10.1007/s00159-014-0078-7} {\bibfield  {journal} {\bibinfo  {journal} {The
  Astronomy and Astrophysics Review}\ }\textbf {\bibinfo {volume} {22}},\
  \bibinfo {pages} {78} (\bibinfo {year} {2014})},\ \Eprint
  {http://arxiv.org/abs/1410.4214} {arXiv:1410.4214} \BibitemShut {NoStop}%
\bibitem [{\citenamefont {Gleeson}\ and\ \citenamefont
  {Axford}(1968)}]{Gleeson:1968zza}%
  \BibitemOpen
  \bibfield  {author} {\bibinfo {author} {\bibfnamefont {L.~J.}\ \bibnamefont
  {Gleeson}}\ and\ \bibinfo {author} {\bibfnamefont {W.~I.}\ \bibnamefont
  {Axford}},\ }\bibfield  {title} {\enquote {\bibinfo {title} {{Solar
  Modulation of Galactic Cosmic Rays}},}\ }\href {\doibase 10.1086/149822}
  {\bibfield  {journal} {\bibinfo  {journal} {Astrophys. J.}\ }\textbf
  {\bibinfo {volume} {154}},\ \bibinfo {pages} {1011} (\bibinfo {year}
  {1968})}\BibitemShut {NoStop}%
\bibitem [{\citenamefont {Cholis}\ \emph {et~al.}(2016)\citenamefont {Cholis},
  \citenamefont {Hooper},\ and\ \citenamefont {Linden}}]{Cholis:2015gna}%
  \BibitemOpen
  \bibfield  {author} {\bibinfo {author} {\bibfnamefont {Ilias}\ \bibnamefont
  {Cholis}}, \bibinfo {author} {\bibfnamefont {Dan}\ \bibnamefont {Hooper}}, \
  and\ \bibinfo {author} {\bibfnamefont {Tim}\ \bibnamefont {Linden}},\
  }\bibfield  {title} {\enquote {\bibinfo {title} {{A Predictive Analytic Model
  for the Solar Modulation of Cosmic Rays}},}\ }\href {\doibase
  10.1103/PhysRevD.93.043016} {\bibfield  {journal} {\bibinfo  {journal} {Phys.
  Rev.}\ }\textbf {\bibinfo {volume} {D93}},\ \bibinfo {pages} {043016}
  (\bibinfo {year} {2016})},\ \Eprint {http://arxiv.org/abs/1511.01507}
  {arXiv:1511.01507 [astro-ph.SR]} \BibitemShut {NoStop}%
\bibitem [{\citenamefont {Bobik}\ \emph {et~al.}(2012)\citenamefont {Bobik}
  \emph {et~al.}}]{Bobik:2011ig}%
  \BibitemOpen
  \bibfield  {author} {\bibinfo {author} {\bibfnamefont {P.}~\bibnamefont
  {Bobik}} \emph {et~al.},\ }\bibfield  {title} {\enquote {\bibinfo {title}
  {{Systematic Investigation of Solar Modulation of Galactic Protons for Solar
  Cycle 23 using a Monte Carlo Approach with Particle Drift Effects and
  Latitudinal Dependence}},}\ }\href {\doibase 10.1088/0004-637X/745/2/132}
  {\bibfield  {journal} {\bibinfo  {journal} {Astrophys. J.}\ }\textbf
  {\bibinfo {volume} {745}},\ \bibinfo {pages} {132} (\bibinfo {year}
  {2012})},\ \Eprint {http://arxiv.org/abs/1110.4315} {arXiv:1110.4315
  [astro-ph.SR]} \BibitemShut {NoStop}%
\bibitem [{\citenamefont {Miyake}\ and\ \citenamefont
  {Yanagita}(2006)}]{Miyake:2006zp}%
  \BibitemOpen
  \bibfield  {author} {\bibinfo {author} {\bibfnamefont {Shoko}\ \bibnamefont
  {Miyake}}\ and\ \bibinfo {author} {\bibfnamefont {Shohei}\ \bibnamefont
  {Yanagita}},\ }\bibfield  {title} {\enquote {\bibinfo {title} {{Effects of
  the tilted and wavy current sheet on the solar modulation of galactic cosmic
  rays}},}\ }in\ \href@noop {} {\emph {\bibinfo {booktitle} {{29th
  International Cosmic Ray Conference (ICRC 2005) Pune, India, August 3-11,
  2005}}}}\ (\bibinfo {year} {2006})\ \Eprint
  {http://arxiv.org/abs/astro-ph/0610777} {arXiv:astro-ph/0610777 [astro-ph]}
  \BibitemShut {NoStop}%
\bibitem [{\citenamefont {Abeysekara}\ \emph {et~al.}(2013)\citenamefont
  {Abeysekara} \emph {et~al.}}]{Abeysekara:2013tza}%
  \BibitemOpen
  \bibfield  {author} {\bibinfo {author} {\bibfnamefont {A.~U.}\ \bibnamefont
  {Abeysekara}} \emph {et~al.},\ }\bibfield  {title} {\enquote {\bibinfo
  {title} {{Sensitivity of the High Altitude Water Cherenkov Detector to
  Sources of Multi-TeV Gamma Rays}},}\ }\href {\doibase
  10.1016/j.astropartphys.2013.08.002} {\bibfield  {journal} {\bibinfo
  {journal} {Astropart. Phys.}\ }\textbf {\bibinfo {volume} {50-52}},\ \bibinfo
  {pages} {26--32} (\bibinfo {year} {2013})},\ \Eprint
  {http://arxiv.org/abs/1306.5800} {arXiv:1306.5800 [astro-ph.HE]} \BibitemShut
  {NoStop}%
\bibitem [{\citenamefont {Ambrosio}\ \emph {et~al.}(1998)\citenamefont
  {Ambrosio} \emph {et~al.}}]{Ambrosio:1997qh}%
  \BibitemOpen
  \bibfield  {author} {\bibinfo {author} {\bibfnamefont {M.}~\bibnamefont
  {Ambrosio}} \emph {et~al.} (\bibinfo {collaboration} {MACRO Collaboration}),\
  }\bibfield  {title} {\enquote {\bibinfo {title} {{The Observation of upgoing
  charged particles produced by high-energy muons in underground detectors}},}\
  }\href {\doibase 10.1016/S0927-6505(98)00010-3} {\bibfield  {journal}
  {\bibinfo  {journal} {Astropart. Phys.}\ }\textbf {\bibinfo {volume} {9}},\
  \bibinfo {pages} {105--117} (\bibinfo {year} {1998})},\ \Eprint
  {http://arxiv.org/abs/hep-ex/9807032} {arXiv:hep-ex/9807032 [hep-ex]}
  \BibitemShut {NoStop}%
\bibitem [{\citenamefont {Kelner}\ \emph {et~al.}(2006)\citenamefont {Kelner},
  \citenamefont {Aharonian},\ and\ \citenamefont {Bugayov}}]{Kelner:2006tc}%
  \BibitemOpen
  \bibfield  {author} {\bibinfo {author} {\bibfnamefont {S.~R.}\ \bibnamefont
  {Kelner}}, \bibinfo {author} {\bibfnamefont {Felex~A.}\ \bibnamefont
  {Aharonian}}, \ and\ \bibinfo {author} {\bibfnamefont {V.~V.}\ \bibnamefont
  {Bugayov}},\ }\bibfield  {title} {\enquote {\bibinfo {title} {{Energy spectra
  of gamma-rays, electrons and neutrinos produced at proton-proton interactions
  in the very high energy regime}},}\ }\href {\doibase
  10.1103/PhysRevD.74.034018, 10.1103/PhysRevD.79.039901} {\bibfield  {journal}
  {\bibinfo  {journal} {Phys. Rev.}\ }\textbf {\bibinfo {volume} {D74}},\
  \bibinfo {pages} {034018} (\bibinfo {year} {2006})},\ \bibinfo {note}
  {[Erratum: Phys. Rev. D79, 039901 (2009)]},\ \Eprint
  {http://arxiv.org/abs/astro-ph/0606058} {arXiv:astro-ph/0606058 [astro-ph]}
  \BibitemShut {NoStop}%
\bibitem [{\citenamefont {Agostinelli}\ \emph {et~al.}(2003)\citenamefont
  {Agostinelli} \emph {et~al.}}]{Agostinelli:2002hh}%
  \BibitemOpen
  \bibfield  {author} {\bibinfo {author} {\bibfnamefont {S.}~\bibnamefont
  {Agostinelli}} \emph {et~al.} (\bibinfo {collaboration} {GEANT4
  Collaboration}),\ }\bibfield  {title} {\enquote {\bibinfo {title} {{GEANT4: A
  Simulation toolkit}},}\ }\href {\doibase 10.1016/S0168-9002(03)01368-8}
  {\bibfield  {journal} {\bibinfo  {journal} {Nucl. Instrum. Meth.}\ }\textbf
  {\bibinfo {volume} {A506}},\ \bibinfo {pages} {250--303} (\bibinfo {year}
  {2003})}\BibitemShut {NoStop}%
\bibitem [{Gea()}]{Geant4}%
  \BibitemOpen
  \href@noop {} {}\bibinfo {howpublished}
  {\url{http://geant4.cern.ch/support/proc_mod_catalog/physics_lists/}}\BibitemShut
  {NoStop}%
\bibitem [{\citenamefont {Baker}\ and\ \citenamefont
  {Temesv{\'a}ry}(1966)}]{baker1966tables}%
  \BibitemOpen
  \bibfield  {author} {\bibinfo {author} {\bibfnamefont {Norman~Hodgson}\
  \bibnamefont {Baker}}\ and\ \bibinfo {author} {\bibfnamefont {Stefan}\
  \bibnamefont {Temesv{\'a}ry}},\ }\href@noop {} {\emph {\bibinfo {title}
  {{Tables of convective stellar envelope models}}}}\ (\bibinfo  {publisher}
  {Institute for Space Studies, Goddard Space Flight Center, National
  Aeronautics and Space Administration},\ \bibinfo {year} {1966})\BibitemShut
  {NoStop}%
\bibitem [{\citenamefont {{Vernazza}}\ \emph {et~al.}(1973)\citenamefont
  {{Vernazza}}, \citenamefont {{Avrett}},\ and\ \citenamefont
  {{Loeser}}}]{1973ApJ...184..605V}%
  \BibitemOpen
  \bibfield  {author} {\bibinfo {author} {\bibfnamefont {J.~E.}\ \bibnamefont
  {{Vernazza}}}, \bibinfo {author} {\bibfnamefont {E.~H.}\ \bibnamefont
  {{Avrett}}}, \ and\ \bibinfo {author} {\bibfnamefont {R.}~\bibnamefont
  {{Loeser}}},\ }\bibfield  {title} {\enquote {\bibinfo {title} {{Structure of
  the Solar Chromosphere. Basic Computations and Summary of the Results}},}\
  }\href {\doibase 10.1086/152353} {\bibfield  {journal} {\bibinfo  {journal}
  {\apj}\ }\textbf {\bibinfo {volume} {184}},\ \bibinfo {pages} {605--632}
  (\bibinfo {year} {1973})}\BibitemShut {NoStop}%
\bibitem [{\citenamefont {Olive}\ \emph {et~al.}(2014)\citenamefont {Olive}
  \emph {et~al.}}]{Agashe:2014kda}%
  \BibitemOpen
  \bibfield  {author} {\bibinfo {author} {\bibfnamefont {K.~A.}\ \bibnamefont
  {Olive}} \emph {et~al.} (\bibinfo {collaboration} {Particle Data Group}),\
  }\bibfield  {title} {\enquote {\bibinfo {title} {{Review of Particle
  Physics}},}\ }\href {\doibase 10.1088/1674-1137/38/9/090001} {\bibfield
  {journal} {\bibinfo  {journal} {Chin. Phys.}\ }\textbf {\bibinfo {volume}
  {C38}},\ \bibinfo {pages} {090001} (\bibinfo {year} {2014})}\BibitemShut
  {NoStop}%
\bibitem [{\citenamefont {Aguilar}\ \emph
  {et~al.}(2015{\natexlab{a}})\citenamefont {Aguilar} \emph
  {et~al.}}]{Aguilar:2015ooa}%
  \BibitemOpen
  \bibfield  {author} {\bibinfo {author} {\bibfnamefont {M.}~\bibnamefont
  {Aguilar}} \emph {et~al.} (\bibinfo {collaboration} {AMS Collaboration}),\
  }\bibfield  {title} {\enquote {\bibinfo {title} {{Precision Measurement of
  the Proton Flux in Primary Cosmic Rays from Rigidity 1 GV to 1.8 TV with the
  Alpha Magnetic Spectrometer on the International Space Station}},}\ }\href
  {\doibase 10.1103/PhysRevLett.114.171103} {\bibfield  {journal} {\bibinfo
  {journal} {Phys. Rev. Lett.}\ }\textbf {\bibinfo {volume} {114}},\ \bibinfo
  {pages} {171103} (\bibinfo {year} {2015}{\natexlab{a}})}\BibitemShut
  {NoStop}%
\bibitem [{\citenamefont {Abdo}\ \emph {et~al.}(2009)\citenamefont {Abdo} \emph
  {et~al.}}]{Abdo:2009gt}%
  \BibitemOpen
  \bibfield  {author} {\bibinfo {author} {\bibfnamefont {A.~A.}\ \bibnamefont
  {Abdo}} \emph {et~al.} (\bibinfo {collaboration} {Fermi-LAT}),\ }\bibfield
  {title} {\enquote {\bibinfo {title} {{Fermi Large Area Telescope Observations
  of the Cosmic-Ray Induced gamma-ray Emission of the Earth's Atmosphere}},}\
  }\href {\doibase 10.1103/PhysRevD.80.122004} {\bibfield  {journal} {\bibinfo
  {journal} {Phys. Rev.}\ }\textbf {\bibinfo {volume} {D80}},\ \bibinfo {pages}
  {122004} (\bibinfo {year} {2009})},\ \Eprint {http://arxiv.org/abs/0912.1868}
  {arXiv:0912.1868 [astro-ph.HE]} \BibitemShut {NoStop}%
\bibitem [{\citenamefont {Ackermann}\ \emph
  {et~al.}(2014{\natexlab{b}})\citenamefont {Ackermann} \emph
  {et~al.}}]{Ackermann:2014ula}%
  \BibitemOpen
  \bibfield  {author} {\bibinfo {author} {\bibfnamefont {M.}~\bibnamefont
  {Ackermann}} \emph {et~al.} (\bibinfo {collaboration} {Fermi-LAT}),\
  }\bibfield  {title} {\enquote {\bibinfo {title} {{Inferred Cosmic-Ray
  Spectrum from Fermi Large Area Telescope γ-Ray Observations of Earth's
  Limb}},}\ }\href {\doibase 10.1103/PhysRevLett.112.151103} {\bibfield
  {journal} {\bibinfo  {journal} {Phys. Rev. Lett.}\ }\textbf {\bibinfo
  {volume} {112}},\ \bibinfo {pages} {151103} (\bibinfo {year}
  {2014}{\natexlab{b}})},\ \Eprint {http://arxiv.org/abs/1403.5372}
  {arXiv:1403.5372 [astro-ph.HE]} \BibitemShut {NoStop}%
\bibitem [{\citenamefont {Rott}\ \emph {et~al.}(2013)\citenamefont {Rott},
  \citenamefont {Siegal-Gaskins},\ and\ \citenamefont {Beacom}}]{Rott:2012qb}%
  \BibitemOpen
  \bibfield  {author} {\bibinfo {author} {\bibfnamefont {Carsten}\ \bibnamefont
  {Rott}}, \bibinfo {author} {\bibfnamefont {Jennifer}\ \bibnamefont
  {Siegal-Gaskins}}, \ and\ \bibinfo {author} {\bibfnamefont {John~F.}\
  \bibnamefont {Beacom}},\ }\bibfield  {title} {\enquote {\bibinfo {title}
  {{New Sensitivity to Solar WIMP Annihilation using Low-Energy Neutrinos}},}\
  }\href {\doibase 10.1103/PhysRevD.88.055005} {\bibfield  {journal} {\bibinfo
  {journal} {Phys. Rev.}\ }\textbf {\bibinfo {volume} {D88}},\ \bibinfo {pages}
  {055005} (\bibinfo {year} {2013})},\ \Eprint {http://arxiv.org/abs/1208.0827}
  {arXiv:1208.0827 [astro-ph.HE]} \BibitemShut {NoStop}%
\bibitem [{\citenamefont {Gaisser}\ and\ \citenamefont
  {Honda}(2002)}]{Gaisser:2002jj}%
  \BibitemOpen
  \bibfield  {author} {\bibinfo {author} {\bibfnamefont {T.~K.}\ \bibnamefont
  {Gaisser}}\ and\ \bibinfo {author} {\bibfnamefont {M.}~\bibnamefont
  {Honda}},\ }\bibfield  {title} {\enquote {\bibinfo {title} {{Flux of
  atmospheric neutrinos}},}\ }\href {\doibase
  10.1146/annurev.nucl.52.050102.090645} {\bibfield  {journal} {\bibinfo
  {journal} {Ann. Rev. Nucl. Part. Sci.}\ }\textbf {\bibinfo {volume} {52}},\
  \bibinfo {pages} {153--199} (\bibinfo {year} {2002})},\ \Eprint
  {http://arxiv.org/abs/hep-ph/0203272} {arXiv:hep-ph/0203272 [hep-ph]}
  \BibitemShut {NoStop}%
\bibitem [{\citenamefont {Aguilar}\ \emph
  {et~al.}(2015{\natexlab{b}})\citenamefont {Aguilar} \emph
  {et~al.}}]{Aguilar:2015ctt}%
  \BibitemOpen
  \bibfield  {author} {\bibinfo {author} {\bibfnamefont {M.}~\bibnamefont
  {Aguilar}} \emph {et~al.} (\bibinfo {collaboration} {AMS Collaboration}),\
  }\bibfield  {title} {\enquote {\bibinfo {title} {{Precision Measurement of
  the Helium Flux in Primary Cosmic Rays of Rigidities 1.9 GV to 3 TV with the
  Alpha Magnetic Spectrometer on the International Space Station}},}\ }\href
  {\doibase 10.1103/PhysRevLett.115.211101} {\bibfield  {journal} {\bibinfo
  {journal} {Phys. Rev. Lett.}\ }\textbf {\bibinfo {volume} {115}},\ \bibinfo
  {pages} {211101} (\bibinfo {year} {2015}{\natexlab{b}})}\BibitemShut
  {NoStop}%
\bibitem [{\citenamefont {Kachelriess}\ \emph {et~al.}(2014)\citenamefont
  {Kachelriess}, \citenamefont {Moskalenko},\ and\ \citenamefont
  {Ostapchenko}}]{Kachelriess:2014mga}%
  \BibitemOpen
  \bibfield  {author} {\bibinfo {author} {\bibfnamefont {Michael}\ \bibnamefont
  {Kachelriess}}, \bibinfo {author} {\bibfnamefont {Igor~V.}\ \bibnamefont
  {Moskalenko}}, \ and\ \bibinfo {author} {\bibfnamefont {Sergey~S.}\
  \bibnamefont {Ostapchenko}},\ }\bibfield  {title} {\enquote {\bibinfo {title}
  {{Nuclear enhancement of the photon yield in cosmic ray interactions}},}\
  }\href {\doibase 10.1088/0004-637X/789/2/136} {\bibfield  {journal} {\bibinfo
   {journal} {Astrophys. J.}\ }\textbf {\bibinfo {volume} {789}},\ \bibinfo
  {pages} {136} (\bibinfo {year} {2014})},\ \Eprint
  {http://arxiv.org/abs/1406.0035} {arXiv:1406.0035 [astro-ph.HE]} \BibitemShut
  {NoStop}%
\bibitem [{\citenamefont {Ahn}\ \emph {et~al.}(2010)\citenamefont {Ahn} \emph
  {et~al.}}]{Ahn:2010gv}%
  \BibitemOpen
  \bibfield  {author} {\bibinfo {author} {\bibfnamefont {H.~S.}\ \bibnamefont
  {Ahn}} \emph {et~al.},\ }\bibfield  {title} {\enquote {\bibinfo {title}
  {{Discrepant hardening observed in cosmic-ray elemental spectra}},}\ }\href
  {\doibase 10.1088/2041-8205/714/1/L89} {\bibfield  {journal} {\bibinfo
  {journal} {Astrophys. J.}\ }\textbf {\bibinfo {volume} {714}},\ \bibinfo
  {pages} {L89--L93} (\bibinfo {year} {2010})},\ \Eprint
  {http://arxiv.org/abs/1004.1123} {arXiv:1004.1123 [astro-ph.HE]} \BibitemShut
  {NoStop}%
\bibitem [{\citenamefont {Yoon}\ \emph {et~al.}(2011)\citenamefont {Yoon} \emph
  {et~al.}}]{Yoon:2011aa}%
  \BibitemOpen
  \bibfield  {author} {\bibinfo {author} {\bibfnamefont {Y.~S.}\ \bibnamefont
  {Yoon}} \emph {et~al.},\ }\bibfield  {title} {\enquote {\bibinfo {title}
  {{Cosmic-Ray Proton and Helium Spectra from the First CREAM Flight}},}\
  }\href {\doibase 10.1088/0004-637X/728/2/122} {\bibfield  {journal} {\bibinfo
   {journal} {Astrophys. J.}\ }\textbf {\bibinfo {volume} {728}},\ \bibinfo
  {pages} {122} (\bibinfo {year} {2011})},\ \Eprint
  {http://arxiv.org/abs/1102.2575} {arXiv:1102.2575 [astro-ph.HE]} \BibitemShut
  {NoStop}%
\bibitem [{\citenamefont {Atkin}\ \emph {et~al.}(2017)\citenamefont {Atkin}
  \emph {et~al.}}]{Atkin:2017vhi}%
  \BibitemOpen
  \bibfield  {author} {\bibinfo {author} {\bibfnamefont {E.}~\bibnamefont
  {Atkin}} \emph {et~al.},\ }\bibfield  {title} {\enquote {\bibinfo {title}
  {{First results of the cosmic ray NUCLEON experiment}},}\ }\href@noop {} {\
  (\bibinfo {year} {2017})},\ \Eprint {http://arxiv.org/abs/1702.02352}
  {arXiv:1702.02352 [astro-ph.HE]} \BibitemShut {NoStop}%
\bibitem [{\citenamefont {Moskalenko}\ \emph {et~al.}(1991)\citenamefont
  {Moskalenko}, \citenamefont {Karakula},\ and\ \citenamefont
  {Tkaczyk}}]{Moskalenko:1991hm}%
  \BibitemOpen
  \bibfield  {author} {\bibinfo {author} {\bibfnamefont {I.~V.}\ \bibnamefont
  {Moskalenko}}, \bibinfo {author} {\bibfnamefont {S.}~\bibnamefont
  {Karakula}}, \ and\ \bibinfo {author} {\bibfnamefont {W.}~\bibnamefont
  {Tkaczyk}},\ }\bibfield  {title} {\enquote {\bibinfo {title} {{The Sun as the
  source of VHE neutrinos}},}\ }\href@noop {} {\bibfield  {journal} {\bibinfo
  {journal} {Astron. Astrophys.}\ }\textbf {\bibinfo {volume} {248}},\ \bibinfo
  {pages} {L5--L6} (\bibinfo {year} {1991})}\BibitemShut {NoStop}%
\bibitem [{\citenamefont {Moskalenko}\ and\ \citenamefont
  {Karakula}(1993)}]{Moskalenko:1993ke}%
  \BibitemOpen
  \bibfield  {author} {\bibinfo {author} {\bibfnamefont {I.~V.}\ \bibnamefont
  {Moskalenko}}\ and\ \bibinfo {author} {\bibfnamefont {S.}~\bibnamefont
  {Karakula}},\ }\bibfield  {title} {\enquote {\bibinfo {title} {{Very
  high-energy neutrinos from the sun}},}\ }\href {\doibase
  10.1088/0954-3899/19/9/019} {\bibfield  {journal} {\bibinfo  {journal} {J.
  Phys.}\ }\textbf {\bibinfo {volume} {G19}},\ \bibinfo {pages} {1399--1406}
  (\bibinfo {year} {1993})}\BibitemShut {NoStop}%
\bibitem [{\citenamefont {Ingelman}\ and\ \citenamefont
  {Thunman}(1996)}]{Ingelman:1996mj}%
  \BibitemOpen
  \bibfield  {author} {\bibinfo {author} {\bibfnamefont {G.}~\bibnamefont
  {Ingelman}}\ and\ \bibinfo {author} {\bibfnamefont {M.}~\bibnamefont
  {Thunman}},\ }\bibfield  {title} {\enquote {\bibinfo {title} {{High-energy
  neutrino production by cosmic ray interactions in the sun}},}\ }\href
  {\doibase 10.1103/PhysRevD.54.4385} {\bibfield  {journal} {\bibinfo
  {journal} {Phys. Rev.}\ }\textbf {\bibinfo {volume} {D54}},\ \bibinfo {pages}
  {4385--4392} (\bibinfo {year} {1996})},\ \Eprint
  {http://arxiv.org/abs/hep-ph/9604288} {arXiv:hep-ph/9604288 [hep-ph]}
  \BibitemShut {NoStop}%
\bibitem [{\citenamefont {Fogli}\ \emph {et~al.}(2006)\citenamefont {Fogli},
  \citenamefont {Lisi}, \citenamefont {Mirizzi}, \citenamefont {Montanino},\
  and\ \citenamefont {Serpico}}]{Fogli:2006jk}%
  \BibitemOpen
  \bibfield  {author} {\bibinfo {author} {\bibfnamefont {G.~L.}\ \bibnamefont
  {Fogli}}, \bibinfo {author} {\bibfnamefont {E.}~\bibnamefont {Lisi}},
  \bibinfo {author} {\bibfnamefont {A.}~\bibnamefont {Mirizzi}}, \bibinfo
  {author} {\bibfnamefont {D.}~\bibnamefont {Montanino}}, \ and\ \bibinfo
  {author} {\bibfnamefont {P.~D.}\ \bibnamefont {Serpico}},\ }\bibfield
  {title} {\enquote {\bibinfo {title} {{Oscillations of solar atmosphere
  neutrinos}},}\ }\href {\doibase 10.1103/PhysRevD.74.093004} {\bibfield
  {journal} {\bibinfo  {journal} {Phys. Rev.}\ }\textbf {\bibinfo {volume}
  {D74}},\ \bibinfo {pages} {093004} (\bibinfo {year} {2006})},\ \Eprint
  {http://arxiv.org/abs/hep-ph/0608321} {arXiv:hep-ph/0608321 [hep-ph]}
  \BibitemShut {NoStop}%
\bibitem [{\citenamefont {Tanaka}\ \emph {et~al.}(2011)\citenamefont {Tanaka}
  \emph {et~al.}}]{Tanaka:2011uf}%
  \BibitemOpen
  \bibfield  {author} {\bibinfo {author} {\bibfnamefont {T.}~\bibnamefont
  {Tanaka}} \emph {et~al.} (\bibinfo {collaboration} {Super-Kamiokande}),\
  }\bibfield  {title} {\enquote {\bibinfo {title} {{An Indirect Search for
  WIMPs in the Sun using 3109.6 days of upward-going muons in
  Super-Kamiokande}},}\ }\href {\doibase 10.1088/0004-637X/742/2/78} {\bibfield
   {journal} {\bibinfo  {journal} {Astrophys. J.}\ }\textbf {\bibinfo {volume}
  {742}},\ \bibinfo {pages} {78} (\bibinfo {year} {2011})},\ \Eprint
  {http://arxiv.org/abs/1108.3384} {arXiv:1108.3384 [astro-ph.HE]} \BibitemShut
  {NoStop}%
\bibitem [{\citenamefont {Aartsen}\ \emph {et~al.}(2013)\citenamefont {Aartsen}
  \emph {et~al.}}]{Aartsen:2012kia}%
  \BibitemOpen
  \bibfield  {author} {\bibinfo {author} {\bibfnamefont {M.~G.}\ \bibnamefont
  {Aartsen}} \emph {et~al.} (\bibinfo {collaboration} {IceCube}),\ }\bibfield
  {title} {\enquote {\bibinfo {title} {{Search for dark matter annihilations in
  the Sun with the 79-string IceCube detector}},}\ }\href {\doibase
  10.1103/PhysRevLett.110.131302} {\bibfield  {journal} {\bibinfo  {journal}
  {Phys. Rev. Lett.}\ }\textbf {\bibinfo {volume} {110}},\ \bibinfo {pages}
  {131302} (\bibinfo {year} {2013})},\ \Eprint {http://arxiv.org/abs/1212.4097}
  {arXiv:1212.4097 [astro-ph.HE]} \BibitemShut {NoStop}%
\bibitem [{\citenamefont {Rott}\ \emph {et~al.}(2011)\citenamefont {Rott},
  \citenamefont {Tanaka},\ and\ \citenamefont {Itow}}]{Rott:2011fh}%
  \BibitemOpen
  \bibfield  {author} {\bibinfo {author} {\bibfnamefont {C.}~\bibnamefont
  {Rott}}, \bibinfo {author} {\bibfnamefont {T.}~\bibnamefont {Tanaka}}, \ and\
  \bibinfo {author} {\bibfnamefont {Y.}~\bibnamefont {Itow}},\ }\bibfield
  {title} {\enquote {\bibinfo {title} {{Enhanced Sensitivity to Dark Matter
  Self-annihilations in the Sun using Neutrino Spectral Information}},}\ }\href
  {\doibase 10.1088/1475-7516/2011/09/029} {\bibfield  {journal} {\bibinfo
  {journal} {JCAP}\ }\textbf {\bibinfo {volume} {1109}},\ \bibinfo {pages}
  {029} (\bibinfo {year} {2011})},\ \Eprint {http://arxiv.org/abs/1107.3182}
  {arXiv:1107.3182 [astro-ph.HE]} \BibitemShut {NoStop}%
\bibitem [{\citenamefont {Argüelles}\ \emph {et~al.}(2017)\citenamefont
  {Argüelles}, \citenamefont {de~Wasseige}, \citenamefont {Fedynitch},\ and\
  \citenamefont {Jones}}]{Arguelles:2017eao}%
  \BibitemOpen
  \bibfield  {author} {\bibinfo {author} {\bibfnamefont {C.~A.}\ \bibnamefont
  {Argüelles}}, \bibinfo {author} {\bibfnamefont {G.}~\bibnamefont
  {de~Wasseige}}, \bibinfo {author} {\bibfnamefont {A.}~\bibnamefont
  {Fedynitch}}, \ and\ \bibinfo {author} {\bibfnamefont {B.~J.~P.}\
  \bibnamefont {Jones}},\ }\bibfield  {title} {\enquote {\bibinfo {title}
  {{Solar Atmospheric Neutrinos and the Sensitivity Floor for Solar Dark Matter
  Annihilation Searches}},}\ }\href@noop {} {\  (\bibinfo {year} {2017})},\
  \Eprint {http://arxiv.org/abs/1703.07798} {arXiv:1703.07798 [astro-ph.HE]}
  \BibitemShut {NoStop}%
\bibitem [{\citenamefont {Ng}\ \emph {et~al.}(2017)\citenamefont {Ng},
  \citenamefont {Beacom}, \citenamefont {Peter},\ and\ \citenamefont
  {Rott}}]{Ng:2017aur}%
  \BibitemOpen
  \bibfield  {author} {\bibinfo {author} {\bibfnamefont {Kenny C.~Y.}\
  \bibnamefont {Ng}}, \bibinfo {author} {\bibfnamefont {John~F.}\ \bibnamefont
  {Beacom}}, \bibinfo {author} {\bibfnamefont {Annika H.~G.}\ \bibnamefont
  {Peter}}, \ and\ \bibinfo {author} {\bibfnamefont {Carsten}\ \bibnamefont
  {Rott}},\ }\bibfield  {title} {\enquote {\bibinfo {title} {{Solar Atmospheric
  Neutrinos: A New Neutrino Floor for Dark Matter Searches}},}\ }\href@noop {}
  {\  (\bibinfo {year} {2017})},\ \Eprint {http://arxiv.org/abs/1703.10280}
  {arXiv:1703.10280 [astro-ph.HE]} \BibitemShut {NoStop}%
\bibitem [{\citenamefont {Edsjo}\ \emph {et~al.}(2017)\citenamefont {Edsjo},
  \citenamefont {Elevant}, \citenamefont {Enberg},\ and\ \citenamefont
  {Niblaeus}}]{Edsjo:2017kjk}%
  \BibitemOpen
  \bibfield  {author} {\bibinfo {author} {\bibfnamefont {Joakim}\ \bibnamefont
  {Edsjo}}, \bibinfo {author} {\bibfnamefont {Jessica}\ \bibnamefont
  {Elevant}}, \bibinfo {author} {\bibfnamefont {Rikard}\ \bibnamefont
  {Enberg}}, \ and\ \bibinfo {author} {\bibfnamefont {Carl}\ \bibnamefont
  {Niblaeus}},\ }\bibfield  {title} {\enquote {\bibinfo {title} {{Neutrinos
  from cosmic ray interactions in the Sun}},}\ }\href@noop {} {\  (\bibinfo
  {year} {2017})},\ \Eprint {http://arxiv.org/abs/1704.02892} {arXiv:1704.02892
  [astro-ph.HE]} \BibitemShut {NoStop}%
\bibitem [{\citenamefont {Ackermann}\ \emph {et~al.}(2012)\citenamefont
  {Ackermann} \emph {et~al.}}]{FermiLAT:2011ab}%
  \BibitemOpen
  \bibfield  {author} {\bibinfo {author} {\bibfnamefont {M.}~\bibnamefont
  {Ackermann}} \emph {et~al.} (\bibinfo {collaboration} {Fermi-LAT
  Collaboration}),\ }\bibfield  {title} {\enquote {\bibinfo {title}
  {{Measurement of separate cosmic-ray electron and positron spectra with the
  Fermi Large Area Telescope}},}\ }\href {\doibase
  10.1103/PhysRevLett.108.011103} {\bibfield  {journal} {\bibinfo  {journal}
  {Phys. Rev. Lett.}\ }\textbf {\bibinfo {volume} {108}},\ \bibinfo {pages}
  {011103} (\bibinfo {year} {2012})},\ \Eprint {http://arxiv.org/abs/1109.0521}
  {arXiv:1109.0521 [astro-ph.HE]} \BibitemShut {NoStop}%
\bibitem [{\citenamefont {Aguilar}\ \emph {et~al.}(2013)\citenamefont {Aguilar}
  \emph {et~al.}}]{Aguilar:2013qda}%
  \BibitemOpen
  \bibfield  {author} {\bibinfo {author} {\bibfnamefont {M.}~\bibnamefont
  {Aguilar}} \emph {et~al.} (\bibinfo {collaboration} {AMS Collaboration}),\
  }\bibfield  {title} {\enquote {\bibinfo {title} {{First Result from the Alpha
  Magnetic Spectrometer on the International Space Station: Precision
  Measurement of the Positron Fraction in Primary Cosmic Rays of 0.5--350
  GeV}},}\ }\href {\doibase 10.1103/PhysRevLett.110.141102} {\bibfield
  {journal} {\bibinfo  {journal} {Phys. Rev. Lett.}\ }\textbf {\bibinfo
  {volume} {110}},\ \bibinfo {pages} {141102} (\bibinfo {year}
  {2013})}\BibitemShut {NoStop}%
\bibitem [{\citenamefont {Torii}(2011)}]{Torii:2011zza}%
  \BibitemOpen
  \bibfield  {author} {\bibinfo {author} {\bibfnamefont {Shoji}\ \bibnamefont
  {Torii}} (\bibinfo {collaboration} {CALET Collaboration}),\ }\bibfield
  {title} {\enquote {\bibinfo {title} {{Calorimetric electron telescope
  mission. Search for dark matter and nearby sources}},}\ }\bibfield
  {booktitle} {\emph {\bibinfo {booktitle} {{Astroparticle physics.
  Proceedings, 2nd Roma International Conference, RICAP'09, Rome, Italy, May
  13-15, 2009}}},\ }\href {\doibase 10.1016/j.nima.2010.06.026} {\bibfield
  {journal} {\bibinfo  {journal} {Nucl. Instrum. Meth.}\ }\textbf {\bibinfo
  {volume} {A630}},\ \bibinfo {pages} {55--57} (\bibinfo {year}
  {2011})}\BibitemShut {NoStop}%
\bibitem [{\citenamefont {Chang}(2014)}]{ChangJin:550}%
  \BibitemOpen
  \bibfield  {author} {\bibinfo {author} {\bibfnamefont {Jin}\ \bibnamefont
  {Chang}},\ }\bibfield  {title} {\enquote {\bibinfo {title} {{Dark Matter
  Particle Explorer: The First Chinese Cosmic Ray and Hard $\gamma$-ray
  Detector in Space}},}\ }\href {\doibase 10.11728/cjss2014.05.550} {\bibfield
  {journal} {\bibinfo  {journal} {Chin. J. Spac. Sci.}\ }\textbf {\bibinfo
  {volume} {34}},\ \bibinfo {eid} {550} (\bibinfo {year} {2014})}\BibitemShut
  {NoStop}%
\bibitem [{\citenamefont {Heitler}(1954)}]{heitler1954quantum}%
  \BibitemOpen
  \bibfield  {author} {\bibinfo {author} {\bibfnamefont {Walter}\ \bibnamefont
  {Heitler}},\ }\href@noop {} {\emph {\bibinfo {title} {{The quantum theory of
  radiation}}}}\ (\bibinfo  {publisher} {Courier Corporation},\ \bibinfo {year}
  {1954})\BibitemShut {NoStop}%
\bibitem [{\citenamefont {Berezinsky}\ \emph {et~al.}(2006)\citenamefont
  {Berezinsky}, \citenamefont {Gazizov},\ and\ \citenamefont
  {Grigorieva}}]{Berezinsky:2002nc}%
  \BibitemOpen
  \bibfield  {author} {\bibinfo {author} {\bibfnamefont {V.}~\bibnamefont
  {Berezinsky}}, \bibinfo {author} {\bibfnamefont {A.~Z.}\ \bibnamefont
  {Gazizov}}, \ and\ \bibinfo {author} {\bibfnamefont {S.~I.}\ \bibnamefont
  {Grigorieva}},\ }\bibfield  {title} {\enquote {\bibinfo {title} {{On
  astrophysical solution to ultrahigh-energy cosmic rays}},}\ }\href {\doibase
  10.1103/PhysRevD.74.043005} {\bibfield  {journal} {\bibinfo  {journal} {Phys.
  Rev.}\ }\textbf {\bibinfo {volume} {D74}},\ \bibinfo {pages} {043005}
  (\bibinfo {year} {2006})},\ \Eprint {http://arxiv.org/abs/hep-ph/0204357}
  {arXiv:hep-ph/0204357 [hep-ph]} \BibitemShut {NoStop}%
\bibitem [{\citenamefont {Andersen}\ and\ \citenamefont
  {Klein}(2011)}]{Andersen:2011dz}%
  \BibitemOpen
  \bibfield  {author} {\bibinfo {author} {\bibfnamefont {Kristoffer~K.}\
  \bibnamefont {Andersen}}\ and\ \bibinfo {author} {\bibfnamefont {Spencer~R.}\
  \bibnamefont {Klein}},\ }\bibfield  {title} {\enquote {\bibinfo {title}
  {{High energy cosmic-ray interactions with particles from the Sun}},}\ }\href
  {\doibase 10.1103/PhysRevD.83.103519} {\bibfield  {journal} {\bibinfo
  {journal} {Phys. Rev.}\ }\textbf {\bibinfo {volume} {D83}},\ \bibinfo {pages}
  {103519} (\bibinfo {year} {2011})},\ \Eprint {http://arxiv.org/abs/1103.5090}
  {arXiv:1103.5090 [astro-ph.HE]} \BibitemShut {NoStop}%
\bibitem [{\citenamefont {Karakula}\ \emph {et~al.}(1994)\citenamefont
  {Karakula}, \citenamefont {Kociolek}, \citenamefont {Moskalenko},\ and\
  \citenamefont {Tkaczyk}}]{Karakula:1994nv}%
  \BibitemOpen
  \bibfield  {author} {\bibinfo {author} {\bibfnamefont {S.}~\bibnamefont
  {Karakula}}, \bibinfo {author} {\bibfnamefont {G.}~\bibnamefont {Kociolek}},
  \bibinfo {author} {\bibfnamefont {I.~V.}\ \bibnamefont {Moskalenko}}, \ and\
  \bibinfo {author} {\bibfnamefont {W.}~\bibnamefont {Tkaczyk}},\ }\bibfield
  {title} {\enquote {\bibinfo {title} {{Gamma-rays from point Galactic
  sources}},}\ }\href {\doibase 10.1086/192001} {\bibfield  {journal} {\bibinfo
   {journal} {Astrophys. J. Suppl.}\ }\textbf {\bibinfo {volume} {92}},\
  \bibinfo {pages} {481--485} (\bibinfo {year} {1994})}\BibitemShut {NoStop}%
\bibitem [{\citenamefont {Anchordoqui}\ \emph
  {et~al.}(2007{\natexlab{a}})\citenamefont {Anchordoqui}, \citenamefont
  {Beacom}, \citenamefont {Goldberg}, \citenamefont {Palomares-Ruiz},\ and\
  \citenamefont {Weiler}}]{Anchordoqui:2006pd}%
  \BibitemOpen
  \bibfield  {author} {\bibinfo {author} {\bibfnamefont {Luis~A.}\ \bibnamefont
  {Anchordoqui}}, \bibinfo {author} {\bibfnamefont {John~F.}\ \bibnamefont
  {Beacom}}, \bibinfo {author} {\bibfnamefont {Haim}\ \bibnamefont {Goldberg}},
  \bibinfo {author} {\bibfnamefont {Sergio}\ \bibnamefont {Palomares-Ruiz}}, \
  and\ \bibinfo {author} {\bibfnamefont {Thomas~J.}\ \bibnamefont {Weiler}},\
  }\bibfield  {title} {\enquote {\bibinfo {title} {{TeV gamma-rays from
  photo-disintegration/de-excitation of cosmic-ray nuclei}},}\ }\href {\doibase
  10.1103/PhysRevLett.98.121101} {\bibfield  {journal} {\bibinfo  {journal}
  {Phys. Rev. Lett.}\ }\textbf {\bibinfo {volume} {98}},\ \bibinfo {pages}
  {121101} (\bibinfo {year} {2007}{\natexlab{a}})},\ \Eprint
  {http://arxiv.org/abs/astro-ph/0611580} {arXiv:astro-ph/0611580 [astro-ph]}
  \BibitemShut {NoStop}%
\bibitem [{\citenamefont {Anchordoqui}\ \emph
  {et~al.}(2007{\natexlab{b}})\citenamefont {Anchordoqui}, \citenamefont
  {Beacom}, \citenamefont {Goldberg}, \citenamefont {Palomares-Ruiz},\ and\
  \citenamefont {Weiler}}]{Anchordoqui:2006pe}%
  \BibitemOpen
  \bibfield  {author} {\bibinfo {author} {\bibfnamefont {Luis~A.}\ \bibnamefont
  {Anchordoqui}}, \bibinfo {author} {\bibfnamefont {John~F.}\ \bibnamefont
  {Beacom}}, \bibinfo {author} {\bibfnamefont {Haim}\ \bibnamefont {Goldberg}},
  \bibinfo {author} {\bibfnamefont {Sergio}\ \bibnamefont {Palomares-Ruiz}}, \
  and\ \bibinfo {author} {\bibfnamefont {Thomas~J.}\ \bibnamefont {Weiler}},\
  }\bibfield  {title} {\enquote {\bibinfo {title} {{TeV $\gamma^-$ rays and
  neutrinos from photo-disintegration of nuclei in Cygnus OB2}},}\ }\href
  {\doibase 10.1103/PhysRevD.75.063001} {\bibfield  {journal} {\bibinfo
  {journal} {Phys. Rev.}\ }\textbf {\bibinfo {volume} {D75}},\ \bibinfo {pages}
  {063001} (\bibinfo {year} {2007}{\natexlab{b}})},\ \Eprint
  {http://arxiv.org/abs/astro-ph/0611581} {arXiv:astro-ph/0611581 [astro-ph]}
  \BibitemShut {NoStop}%
\bibitem [{\citenamefont {Murase}\ and\ \citenamefont
  {Beacom}(2010)}]{Murase:2010va}%
  \BibitemOpen
  \bibfield  {author} {\bibinfo {author} {\bibfnamefont {Kohta}\ \bibnamefont
  {Murase}}\ and\ \bibinfo {author} {\bibfnamefont {John~F.}\ \bibnamefont
  {Beacom}},\ }\bibfield  {title} {\enquote {\bibinfo {title}
  {{Very-High-Energy Gamma-Ray Signal from Nuclear Photodisintegration as a
  Probe of Extragalactic Sources of Ultrahigh-Energy Nuclei}},}\ }\href
  {\doibase 10.1103/PhysRevD.82.043008} {\bibfield  {journal} {\bibinfo
  {journal} {Phys. Rev.}\ }\textbf {\bibinfo {volume} {D82}},\ \bibinfo {pages}
  {043008} (\bibinfo {year} {2010})},\ \Eprint {http://arxiv.org/abs/1002.3980}
  {arXiv:1002.3980 [astro-ph.HE]} \BibitemShut {NoStop}%
\bibitem [{\citenamefont {Moskalenko}\ and\ \citenamefont
  {Strong}(2000)}]{Moskalenko2000a}%
  \BibitemOpen
  \bibfield  {author} {\bibinfo {author} {\bibfnamefont {Igor~V.}\ \bibnamefont
  {Moskalenko}}\ and\ \bibinfo {author} {\bibfnamefont {Andrew~W.}\
  \bibnamefont {Strong}},\ }\bibfield  {title} {\enquote {\bibinfo {title}
  {{Anisotropic Inverse Compton Scattering in the Galaxy}},}\ }\href {\doibase
  10.1086/308138} {\bibfield  {journal} {\bibinfo  {journal} {The Astrophysical
  Journal}\ }\textbf {\bibinfo {volume} {528}},\ \bibinfo {pages} {357--367}
  (\bibinfo {year} {2000})},\ \Eprint {http://arxiv.org/abs/9811284}
  {arXiv:9811284 [astro-ph]} \BibitemShut {NoStop}%
\bibitem [{\citenamefont {Orlando}\ and\ \citenamefont
  {Strong}(2013{\natexlab{a}})}]{Orlando:2013pza}%
  \BibitemOpen
  \bibfield  {author} {\bibinfo {author} {\bibfnamefont {Elena}\ \bibnamefont
  {Orlando}}\ and\ \bibinfo {author} {\bibfnamefont {Andrew}\ \bibnamefont
  {Strong}},\ }\bibfield  {title} {\enquote {\bibinfo {title} {{StellarICs:
  Stellar and solar Inverse Compton emission package}},}\ }\href@noop {} {\
  (\bibinfo {year} {2013}{\natexlab{a}})},\ \Eprint
  {http://arxiv.org/abs/1307.6798} {arXiv:1307.6798 [astro-ph.HE]} \BibitemShut
  {NoStop}%
\bibitem [{\citenamefont {Orlando}\ and\ \citenamefont
  {Strong}(2013{\natexlab{b}})}]{Orlando:2013nga}%
  \BibitemOpen
  \bibfield  {author} {\bibinfo {author} {\bibfnamefont {Elena}\ \bibnamefont
  {Orlando}}\ and\ \bibinfo {author} {\bibfnamefont {Andrew~W.}\ \bibnamefont
  {Strong}},\ }\bibfield  {title} {\enquote {\bibinfo {title} {{A software
  package for stellar and solar inverse-Compton emission: Stellarics}},}\
  }\bibfield  {booktitle} {\emph {\bibinfo {booktitle} {{Proceedings, 9th
  Workshop on Science with the New Generation of High Energy Gamma-ray
  Experiments (SciNeGHE 2012): Lecce, Italy, June 20-22, 2012}}},\ }\href
  {\doibase 10.1016/j.nuclphysbps.2013.05.042} {\bibfield  {journal} {\bibinfo
  {journal} {Nucl. Phys. Proc. Suppl.}\ }\textbf {\bibinfo {volume}
  {239-240}},\ \bibinfo {pages} {266--269} (\bibinfo {year}
  {2013}{\natexlab{b}})},\ \Eprint {http://arxiv.org/abs/1303.5491}
  {arXiv:1303.5491 [astro-ph.SR]} \BibitemShut {NoStop}%
\bibitem [{\citenamefont {Aguilar}\ \emph {et~al.}(2014)\citenamefont {Aguilar}
  \emph {et~al.}}]{Aguilar:2014mma}%
  \BibitemOpen
  \bibfield  {author} {\bibinfo {author} {\bibfnamefont {M.}~\bibnamefont
  {Aguilar}} \emph {et~al.} (\bibinfo {collaboration} {AMS Collaboration}),\
  }\bibfield  {title} {\enquote {\bibinfo {title} {{Electron and Positron
  Fluxes in Primary Cosmic Rays Measured with the Alpha Magnetic Spectrometer
  on the International Space Station}},}\ }\href {\doibase
  10.1103/PhysRevLett.113.121102} {\bibfield  {journal} {\bibinfo  {journal}
  {Phys. Rev. Lett.}\ }\textbf {\bibinfo {volume} {113}},\ \bibinfo {pages}
  {121102} (\bibinfo {year} {2014})}\BibitemShut {NoStop}%
\bibitem [{\citenamefont {Aharonian}\ \emph {et~al.}(2008)\citenamefont
  {Aharonian} \emph {et~al.}}]{Aharonian:2008aa}%
  \BibitemOpen
  \bibfield  {author} {\bibinfo {author} {\bibfnamefont {F.}~\bibnamefont
  {Aharonian}} \emph {et~al.} (\bibinfo {collaboration} {H.E.S.S.
  Collaboration}),\ }\bibfield  {title} {\enquote {\bibinfo {title} {{The
  energy spectrum of cosmic-ray electrons at TeV energies}},}\ }\href {\doibase
  10.1103/PhysRevLett.101.261104} {\bibfield  {journal} {\bibinfo  {journal}
  {Phys. Rev. Lett.}\ }\textbf {\bibinfo {volume} {101}},\ \bibinfo {pages}
  {261104} (\bibinfo {year} {2008})},\ \Eprint {http://arxiv.org/abs/0811.3894}
  {arXiv:0811.3894 [astro-ph]} \BibitemShut {NoStop}%
\bibitem [{\citenamefont {Aharonian}\ \emph {et~al.}(2009)\citenamefont
  {Aharonian} \emph {et~al.}}]{Aharonian:2009ah}%
  \BibitemOpen
  \bibfield  {author} {\bibinfo {author} {\bibfnamefont {F.}~\bibnamefont
  {Aharonian}} \emph {et~al.} (\bibinfo {collaboration} {H.E.S.S.
  Collaboration}),\ }\bibfield  {title} {\enquote {\bibinfo {title} {{Probing
  the ATIC peak in the cosmic-ray electron spectrum with H.E.S.S}},}\ }\href
  {\doibase 10.1051/0004-6361/200913323} {\bibfield  {journal} {\bibinfo
  {journal} {Astron. Astrophys.}\ }\textbf {\bibinfo {volume} {508}},\ \bibinfo
  {pages} {561} (\bibinfo {year} {2009})},\ \Eprint
  {http://arxiv.org/abs/0905.0105} {arXiv:0905.0105 [astro-ph.HE]} \BibitemShut
  {NoStop}%
\bibitem [{\citenamefont {Staszak}(2016)}]{Staszak:2015kza}%
  \BibitemOpen
  \bibfield  {author} {\bibinfo {author} {\bibfnamefont {David}\ \bibnamefont
  {Staszak}} (\bibinfo {collaboration} {VERITAS Collaboration}),\ }\bibfield
  {title} {\enquote {\bibinfo {title} {{A Cosmic-ray Electron Spectrum with
  VERITAS}},}\ }\bibfield  {booktitle} {\emph {\bibinfo {booktitle}
  {{Proceedings, 34th International Cosmic Ray Conference (ICRC 2015): The
  Hague, The Netherlands, July 30-August 6, 2015}}},\ }\href@noop {} {\bibfield
   {journal} {\bibinfo  {journal} {PoS}\ }\textbf {\bibinfo {volume}
  {ICRC2015}},\ \bibinfo {pages} {411} (\bibinfo {year} {2016})},\ \Eprint
  {http://arxiv.org/abs/1508.06597} {arXiv:1508.06597 [astro-ph.HE]}
  \BibitemShut {NoStop}%
\bibitem [{\citenamefont {Kistler}\ and\ \citenamefont
  {Yuksel}(2009)}]{Kistler:2009wm}%
  \BibitemOpen
  \bibfield  {author} {\bibinfo {author} {\bibfnamefont {Matthew~D.}\
  \bibnamefont {Kistler}}\ and\ \bibinfo {author} {\bibfnamefont {Hasan}\
  \bibnamefont {Yuksel}},\ }\bibfield  {title} {\enquote {\bibinfo {title}
  {{New Constraints on the Highest-Energy Cosmic-Ray Electrons and
  Positrons}},}\ }\href@noop {} {\  (\bibinfo {year} {2009})},\ \Eprint
  {http://arxiv.org/abs/0912.0264} {arXiv:0912.0264 [astro-ph.HE]} \BibitemShut
  {NoStop}%
\bibitem [{\citenamefont {Ahlers}\ and\ \citenamefont
  {Murase}(2014)}]{Ahlers:2013xia}%
  \BibitemOpen
  \bibfield  {author} {\bibinfo {author} {\bibfnamefont {Markus}\ \bibnamefont
  {Ahlers}}\ and\ \bibinfo {author} {\bibfnamefont {Kohta}\ \bibnamefont
  {Murase}},\ }\bibfield  {title} {\enquote {\bibinfo {title} {{Probing the
  Galactic Origin of the IceCube Excess with Gamma-Rays}},}\ }\href {\doibase
  10.1103/PhysRevD.90.023010} {\bibfield  {journal} {\bibinfo  {journal} {Phys.
  Rev.}\ }\textbf {\bibinfo {volume} {D90}},\ \bibinfo {pages} {023010}
  (\bibinfo {year} {2014})},\ \Eprint {http://arxiv.org/abs/1309.4077}
  {arXiv:1309.4077 [astro-ph.HE]} \BibitemShut {NoStop}%
\bibitem [{\citenamefont {Chang}\ \emph {et~al.}(2008)\citenamefont {Chang}
  \emph {et~al.}}]{Chang:2008aa}%
  \BibitemOpen
  \bibfield  {author} {\bibinfo {author} {\bibfnamefont {J.}~\bibnamefont
  {Chang}} \emph {et~al.},\ }\bibfield  {title} {\enquote {\bibinfo {title}
  {{An excess of cosmic ray electrons at energies of 300-800 GeV}},}\ }\href
  {\doibase 10.1038/nature07477} {\bibfield  {journal} {\bibinfo  {journal}
  {Nature}\ }\textbf {\bibinfo {volume} {456}},\ \bibinfo {pages} {362--365}
  (\bibinfo {year} {2008})}\BibitemShut {NoStop}%
\bibitem [{\citenamefont {Adriani}\ \emph {et~al.}(2009)\citenamefont {Adriani}
  \emph {et~al.}}]{Adriani:2008zr}%
  \BibitemOpen
  \bibfield  {author} {\bibinfo {author} {\bibfnamefont {Oscar}\ \bibnamefont
  {Adriani}} \emph {et~al.} (\bibinfo {collaboration} {PAMELA Collaboration}),\
  }\bibfield  {title} {\enquote {\bibinfo {title} {{An anomalous positron
  abundance in cosmic rays with energies 1.5-100 GeV}},}\ }\href {\doibase
  10.1038/nature07942} {\bibfield  {journal} {\bibinfo  {journal} {Nature}\
  }\textbf {\bibinfo {volume} {458}},\ \bibinfo {pages} {607--609} (\bibinfo
  {year} {2009})},\ \Eprint {http://arxiv.org/abs/0810.4995} {arXiv:0810.4995
  [astro-ph]} \BibitemShut {NoStop}%
\bibitem [{\citenamefont {Profumo}(2011)}]{Profumo:2008ms}%
  \BibitemOpen
  \bibfield  {author} {\bibinfo {author} {\bibfnamefont {Stefano}\ \bibnamefont
  {Profumo}},\ }\bibfield  {title} {\enquote {\bibinfo {title} {{Dissecting
  cosmic-ray electron-positron data with Occam's Razor: the role of known
  Pulsars}},}\ }\href {\doibase 10.2478/s11534-011-0099-z} {\bibfield
  {journal} {\bibinfo  {journal} {Central Eur. J. Phys.}\ }\textbf {\bibinfo
  {volume} {10}},\ \bibinfo {pages} {1--31} (\bibinfo {year} {2011})},\ \Eprint
  {http://arxiv.org/abs/0812.4457} {arXiv:0812.4457 [astro-ph]} \BibitemShut
  {NoStop}%
\bibitem [{\citenamefont {Hinton}\ \emph {et~al.}(2011)\citenamefont {Hinton},
  \citenamefont {Funk}, \citenamefont {Parsons},\ and\ \citenamefont
  {Ohm}}]{Hinton:2011ad}%
  \BibitemOpen
  \bibfield  {author} {\bibinfo {author} {\bibfnamefont {J.}~\bibnamefont
  {Hinton}}, \bibinfo {author} {\bibfnamefont {S.}~\bibnamefont {Funk}},
  \bibinfo {author} {\bibfnamefont {R.~D.}\ \bibnamefont {Parsons}}, \ and\
  \bibinfo {author} {\bibfnamefont {S.}~\bibnamefont {Ohm}},\ }\bibfield
  {title} {\enquote {\bibinfo {title} {{Escape from Vela X}},}\ }\href
  {\doibase 10.1088/2041-8205/743/1/L7} {\bibfield  {journal} {\bibinfo
  {journal} {Astrophys. J.}\ }\textbf {\bibinfo {volume} {743}},\ \bibinfo
  {pages} {L7} (\bibinfo {year} {2011})},\ \Eprint
  {http://arxiv.org/abs/1111.2036} {arXiv:1111.2036 [astro-ph.HE]} \BibitemShut
  {NoStop}%
\bibitem [{\citenamefont {Adriani}\ \emph {et~al.}(2011)\citenamefont {Adriani}
  \emph {et~al.}}]{Adriani:2011xv}%
  \BibitemOpen
  \bibfield  {author} {\bibinfo {author} {\bibfnamefont {O.}~\bibnamefont
  {Adriani}} \emph {et~al.} (\bibinfo {collaboration} {PAMELA Collaboration}),\
  }\bibfield  {title} {\enquote {\bibinfo {title} {{The cosmic-ray electron
  flux measured by the PAMELA experiment between 1 and 625 GeV}},}\ }\href
  {\doibase 10.1103/PhysRevLett.106.201101} {\bibfield  {journal} {\bibinfo
  {journal} {Phys. Rev. Lett.}\ }\textbf {\bibinfo {volume} {106}},\ \bibinfo
  {pages} {201101} (\bibinfo {year} {2011})},\ \Eprint
  {http://arxiv.org/abs/1103.2880} {arXiv:1103.2880 [astro-ph.HE]} \BibitemShut
  {NoStop}%
\bibitem [{\citenamefont {BenZvi}\ \emph {et~al.}(2016)\citenamefont {BenZvi},
  \citenamefont {Fiorino}, \citenamefont {Hampel-Arias},\ and\ \citenamefont
  {Nisa}}]{BenZvi:2015kga}%
  \BibitemOpen
  \bibfield  {author} {\bibinfo {author} {\bibfnamefont {Segev}\ \bibnamefont
  {BenZvi}}, \bibinfo {author} {\bibfnamefont {Daniel}\ \bibnamefont
  {Fiorino}}, \bibinfo {author} {\bibfnamefont {Zigfried}\ \bibnamefont
  {Hampel-Arias}}, \ and\ \bibinfo {author} {\bibfnamefont {Mehr~Un}\
  \bibnamefont {Nisa}} (\bibinfo {collaboration} {HAWC Collaboration}),\
  }\bibfield  {title} {\enquote {\bibinfo {title} {{Towards a Measurement of
  the $e^+e^-$ Flux above 1 TeV with HAWC}},}\ }\bibfield  {booktitle} {\emph
  {\bibinfo {booktitle} {{Proceedings, 34th International Cosmic Ray Conference
  (ICRC 2015): The Hague, The Netherlands, July 30-August 6, 2015}}},\
  }\href@noop {} {\bibfield  {journal} {\bibinfo  {journal} {PoS}\ }\textbf
  {\bibinfo {volume} {ICRC2015}},\ \bibinfo {pages} {248} (\bibinfo {year}
  {2016})},\ \Eprint {http://arxiv.org/abs/1508.03466} {arXiv:1508.03466
  [astro-ph.HE]} \BibitemShut {NoStop}%
\bibitem [{\citenamefont {Prodanovic}\ \emph {et~al.}(2007)\citenamefont
  {Prodanovic}, \citenamefont {Fields},\ and\ \citenamefont
  {Beacom}}]{Prodanovic:2006bq}%
  \BibitemOpen
  \bibfield  {author} {\bibinfo {author} {\bibfnamefont {Tijana}\ \bibnamefont
  {Prodanovic}}, \bibinfo {author} {\bibfnamefont {Brian~D.}\ \bibnamefont
  {Fields}}, \ and\ \bibinfo {author} {\bibfnamefont {John~F.}\ \bibnamefont
  {Beacom}},\ }\bibfield  {title} {\enquote {\bibinfo {title} {{Diffuse gamma
  rays from the galactic plane: probing the gev excess and identifying the TeV
  excess}},}\ }\href {\doibase 10.1016/j.astropartphys.2006.08.007} {\bibfield
  {journal} {\bibinfo  {journal} {Astropart. Phys.}\ }\textbf {\bibinfo
  {volume} {27}},\ \bibinfo {pages} {10--20} (\bibinfo {year} {2007})},\
  \Eprint {http://arxiv.org/abs/astro-ph/0603618} {arXiv:astro-ph/0603618
  [astro-ph]} \BibitemShut {NoStop}%
\bibitem [{\citenamefont {Ayala~Solares}\ \emph {et~al.}(2016)\citenamefont
  {Ayala~Solares}, \citenamefont {Hui},\ and\ \citenamefont
  {Hüntemeyer}}]{Solares:2015jza}%
  \BibitemOpen
  \bibfield  {author} {\bibinfo {author} {\bibfnamefont {Hugo}\ \bibnamefont
  {Ayala~Solares}}, \bibinfo {author} {\bibfnamefont {C.~Michelle}\
  \bibnamefont {Hui}}, \ and\ \bibinfo {author} {\bibfnamefont {Petra}\
  \bibnamefont {Hüntemeyer}} (\bibinfo {collaboration} {HAWC}),\ }\bibfield
  {title} {\enquote {\bibinfo {title} {{Fermi Bubbles with HAWC}},}\ }\bibfield
   {booktitle} {\emph {\bibinfo {booktitle} {{Proceedings, 34th International
  Cosmic Ray Conference (ICRC 2015): The Hague, The Netherlands, July 30-August
  6, 2015}}},\ }\href@noop {} {\bibfield  {journal} {\bibinfo  {journal} {PoS}\
  }\textbf {\bibinfo {volume} {ICRC2015}},\ \bibinfo {pages} {749} (\bibinfo
  {year} {2016})},\ \Eprint {http://arxiv.org/abs/1508.06592} {arXiv:1508.06592
  [astro-ph.HE]} \BibitemShut {NoStop}%
\bibitem [{\citenamefont {Pretz}(2016{\natexlab{a}})}]{Pretz:2015zja}%
  \BibitemOpen
  \bibfield  {author} {\bibinfo {author} {\bibfnamefont {John}\ \bibnamefont
  {Pretz}} (\bibinfo {collaboration} {HAWC}),\ }\bibfield  {title} {\enquote
  {\bibinfo {title} {{Highlights from the High Altitude Water Cherenkov
  Observatory}},}\ }\bibfield  {booktitle} {\emph {\bibinfo {booktitle}
  {{Proceedings, 34th International Cosmic Ray Conference (ICRC 2015): The
  Hague, The Netherlands, July 30-August 6, 2015}}},\ }\href@noop {} {\bibfield
   {journal} {\bibinfo  {journal} {PoS}\ }\textbf {\bibinfo {volume}
  {ICRC2015}},\ \bibinfo {pages} {025} (\bibinfo {year}
  {2016}{\natexlab{a}})},\ \Eprint {http://arxiv.org/abs/1509.07851}
  {arXiv:1509.07851 [astro-ph.HE]} \BibitemShut {NoStop}%
\bibitem [{\citenamefont {Pretz}(2016{\natexlab{b}})}]{Pretz:2015wma}%
  \BibitemOpen
  \bibfield  {author} {\bibinfo {author} {\bibfnamefont {John}\ \bibnamefont
  {Pretz}} (\bibinfo {collaboration} {HAWC}),\ }\bibfield  {title} {\enquote
  {\bibinfo {title} {{Limit on an Isotropic Diffuse Gamma-Ray Population with
  HAWC}},}\ }\bibfield  {booktitle} {\emph {\bibinfo {booktitle} {{Proceedings,
  34th International Cosmic Ray Conference (ICRC 2015): The Hague, The
  Netherlands, July 30-August 6, 2015}}},\ }\href@noop {} {\bibfield  {journal}
  {\bibinfo  {journal} {PoS}\ }\textbf {\bibinfo {volume} {ICRC2015}},\
  \bibinfo {pages} {820} (\bibinfo {year} {2016}{\natexlab{b}})},\ \Eprint
  {http://arxiv.org/abs/1508.04091} {arXiv:1508.04091 [astro-ph.HE]}
  \BibitemShut {NoStop}%
\bibitem [{\citenamefont {Fang}\ \emph {et~al.}(2016)\citenamefont {Fang},
  \citenamefont {Wang}, \citenamefont {Bi}, \citenamefont {Lin},\ and\
  \citenamefont {Yin}}]{Fang:2016wid}%
  \BibitemOpen
  \bibfield  {author} {\bibinfo {author} {\bibfnamefont {Kun}\ \bibnamefont
  {Fang}}, \bibinfo {author} {\bibfnamefont {Bing-Bing}\ \bibnamefont {Wang}},
  \bibinfo {author} {\bibfnamefont {Xiao-Jun}\ \bibnamefont {Bi}}, \bibinfo
  {author} {\bibfnamefont {Su-Jie}\ \bibnamefont {Lin}}, \ and\ \bibinfo
  {author} {\bibfnamefont {Peng-Fei}\ \bibnamefont {Yin}},\ }\bibfield  {title}
  {\enquote {\bibinfo {title} {{Perspective on the cosmic-ray electron spectrum
  above TeV}},}\ }\href@noop {} {\  (\bibinfo {year} {2016})},\ \Eprint
  {http://arxiv.org/abs/1611.10292} {arXiv:1611.10292 [astro-ph.HE]}
  \BibitemShut {NoStop}%
\bibitem [{\citenamefont {Cirelli}\ \emph {et~al.}(2011)\citenamefont
  {Cirelli}, \citenamefont {Corcella}, \citenamefont {Hektor}, \citenamefont
  {Hutsi}, \citenamefont {Kadastik}, \citenamefont {Panci}, \citenamefont
  {Raidal}, \citenamefont {Sala},\ and\ \citenamefont
  {Strumia}}]{Cirelli:2010xx}%
  \BibitemOpen
  \bibfield  {author} {\bibinfo {author} {\bibfnamefont {Marco}\ \bibnamefont
  {Cirelli}}, \bibinfo {author} {\bibfnamefont {Gennaro}\ \bibnamefont
  {Corcella}}, \bibinfo {author} {\bibfnamefont {Andi}\ \bibnamefont {Hektor}},
  \bibinfo {author} {\bibfnamefont {Gert}\ \bibnamefont {Hutsi}}, \bibinfo
  {author} {\bibfnamefont {Mario}\ \bibnamefont {Kadastik}}, \bibinfo {author}
  {\bibfnamefont {Paolo}\ \bibnamefont {Panci}}, \bibinfo {author}
  {\bibfnamefont {Martti}\ \bibnamefont {Raidal}}, \bibinfo {author}
  {\bibfnamefont {Filippo}\ \bibnamefont {Sala}}, \ and\ \bibinfo {author}
  {\bibfnamefont {Alessandro}\ \bibnamefont {Strumia}},\ }\bibfield  {title}
  {\enquote {\bibinfo {title} {{PPPC 4 DM ID: A Poor Particle Physicist
  Cookbook for Dark Matter Indirect Detection}},}\ }\href {\doibase
  10.1088/1475-7516/2012/10/E01, 10.1088/1475-7516/2011/03/051} {\bibfield
  {journal} {\bibinfo  {journal} {JCAP}\ }\textbf {\bibinfo {volume} {1103}},\
  \bibinfo {pages} {051} (\bibinfo {year} {2011})},\ \bibinfo {note} {[Erratum:
  JCAP1210,E01(2012)]},\ \Eprint {http://arxiv.org/abs/1012.4515}
  {arXiv:1012.4515 [hep-ph]} \BibitemShut {NoStop}%
\bibitem [{\citenamefont {Buch}\ \emph {et~al.}(2015)\citenamefont {Buch},
  \citenamefont {Cirelli}, \citenamefont {Giesen},\ and\ \citenamefont
  {Taoso}}]{Buch:2015iya}%
  \BibitemOpen
  \bibfield  {author} {\bibinfo {author} {\bibfnamefont {Jatan}\ \bibnamefont
  {Buch}}, \bibinfo {author} {\bibfnamefont {Marco}\ \bibnamefont {Cirelli}},
  \bibinfo {author} {\bibfnamefont {Gaëlle}\ \bibnamefont {Giesen}}, \ and\
  \bibinfo {author} {\bibfnamefont {Marco}\ \bibnamefont {Taoso}},\ }\bibfield
  {title} {\enquote {\bibinfo {title} {{PPPC 4 DM secondary: A Poor Particle
  Physicist Cookbook for secondary radiation from Dark Matter}},}\ }\href
  {\doibase 10.1088/1475-7516/2015/9/037, 10.1088/1475-7516/2015/09/037}
  {\bibfield  {journal} {\bibinfo  {journal} {JCAP}\ }\textbf {\bibinfo
  {volume} {1509}},\ \bibinfo {pages} {037} (\bibinfo {year} {2015})},\ \Eprint
  {http://arxiv.org/abs/1505.01049} {arXiv:1505.01049 [hep-ph]} \BibitemShut
  {NoStop}%
\bibitem [{\citenamefont {Murase}\ and\ \citenamefont
  {Beacom}(2012)}]{Murase:2012xs}%
  \BibitemOpen
  \bibfield  {author} {\bibinfo {author} {\bibfnamefont {Kohta}\ \bibnamefont
  {Murase}}\ and\ \bibinfo {author} {\bibfnamefont {John~F.}\ \bibnamefont
  {Beacom}},\ }\bibfield  {title} {\enquote {\bibinfo {title} {{Constraining
  Very Heavy Dark Matter Using Diffuse Backgrounds of Neutrinos and Cascaded
  Gamma Rays}},}\ }\href {\doibase 10.1088/1475-7516/2012/10/043} {\bibfield
  {journal} {\bibinfo  {journal} {JCAP}\ }\textbf {\bibinfo {volume} {1210}},\
  \bibinfo {pages} {043} (\bibinfo {year} {2012})},\ \Eprint
  {http://arxiv.org/abs/1206.2595} {arXiv:1206.2595 [hep-ph]} \BibitemShut
  {NoStop}%
\bibitem [{\citenamefont {Esmaili}\ and\ \citenamefont
  {Serpico}(2013)}]{Esmaili:2013gha}%
  \BibitemOpen
  \bibfield  {author} {\bibinfo {author} {\bibfnamefont {Arman}\ \bibnamefont
  {Esmaili}}\ and\ \bibinfo {author} {\bibfnamefont {Pasquale~Dario}\
  \bibnamefont {Serpico}},\ }\bibfield  {title} {\enquote {\bibinfo {title}
  {{Are IceCube neutrinos unveiling PeV-scale decaying dark matter?}}}\ }\href
  {\doibase 10.1088/1475-7516/2013/11/054} {\bibfield  {journal} {\bibinfo
  {journal} {JCAP}\ }\textbf {\bibinfo {volume} {1311}},\ \bibinfo {pages}
  {054} (\bibinfo {year} {2013})},\ \Eprint {http://arxiv.org/abs/1308.1105}
  {arXiv:1308.1105 [hep-ph]} \BibitemShut {NoStop}%
\bibitem [{\citenamefont {Feldstein}\ \emph {et~al.}(2013)\citenamefont
  {Feldstein}, \citenamefont {Kusenko}, \citenamefont {Matsumoto},\ and\
  \citenamefont {Yanagida}}]{Feldstein:2013kka}%
  \BibitemOpen
  \bibfield  {author} {\bibinfo {author} {\bibfnamefont {Brian}\ \bibnamefont
  {Feldstein}}, \bibinfo {author} {\bibfnamefont {Alexander}\ \bibnamefont
  {Kusenko}}, \bibinfo {author} {\bibfnamefont {Shigeki}\ \bibnamefont
  {Matsumoto}}, \ and\ \bibinfo {author} {\bibfnamefont {Tsutomu~T.}\
  \bibnamefont {Yanagida}},\ }\bibfield  {title} {\enquote {\bibinfo {title}
  {{Neutrinos at IceCube from Heavy Decaying Dark Matter}},}\ }\href {\doibase
  10.1103/PhysRevD.88.015004} {\bibfield  {journal} {\bibinfo  {journal} {Phys.
  Rev.}\ }\textbf {\bibinfo {volume} {D88}},\ \bibinfo {pages} {015004}
  (\bibinfo {year} {2013})},\ \Eprint {http://arxiv.org/abs/1303.7320}
  {arXiv:1303.7320 [hep-ph]} \BibitemShut {NoStop}%
\bibitem [{\citenamefont {Actis}\ \emph {et~al.}(2011)\citenamefont {Actis}
  \emph {et~al.}}]{Consortium:2010bc}%
  \BibitemOpen
  \bibfield  {author} {\bibinfo {author} {\bibfnamefont {M.}~\bibnamefont
  {Actis}} \emph {et~al.} (\bibinfo {collaboration} {CTA Consortium}),\
  }\bibfield  {title} {\enquote {\bibinfo {title} {{Design concepts for the
  Cherenkov Telescope Array CTA: An advanced facility for ground-based
  high-energy gamma-ray astronomy}},}\ }\href {\doibase
  10.1007/s10686-011-9247-0} {\bibfield  {journal} {\bibinfo  {journal} {Exper.
  Astron.}\ }\textbf {\bibinfo {volume} {32}},\ \bibinfo {pages} {193--316}
  (\bibinfo {year} {2011})},\ \Eprint {http://arxiv.org/abs/1008.3703}
  {arXiv:1008.3703 [astro-ph.IM]} \BibitemShut {NoStop}%
\bibitem [{zhe()}]{zheli}%
  \BibitemOpen
  \href@noop {} {}\bibinfo {howpublished}
  {\url{https://agenda.infn.it/contributionDisplay.py?contribId=24&confId=12038}}\BibitemShut
  {NoStop}%
\bibitem [{\citenamefont {Leane}\ \emph {et~al.}(2017)\citenamefont {Leane},
  \citenamefont {Ng},\ and\ \citenamefont {Beacom}}]{Leane:2017vag}%
  \BibitemOpen
  \bibfield  {author} {\bibinfo {author} {\bibfnamefont {Rebecca~K.}\
  \bibnamefont {Leane}}, \bibinfo {author} {\bibfnamefont {Kenny C.~Y.}\
  \bibnamefont {Ng}}, \ and\ \bibinfo {author} {\bibfnamefont {John~F.}\
  \bibnamefont {Beacom}},\ }\bibfield  {title} {\enquote {\bibinfo {title}
  {{Powerful Solar Signatures of Long-Lived Dark Mediators}},}\ }\href@noop {}
  {\  (\bibinfo {year} {2017})},\ \Eprint {http://arxiv.org/abs/1703.04629}
  {arXiv:1703.04629 [astro-ph.HE]} \BibitemShut {NoStop}%
\bibitem [{\citenamefont {Arina}\ \emph {et~al.}(2017)\citenamefont {Arina},
  \citenamefont {Backovic}, \citenamefont {Heisig},\ and\ \citenamefont
  {Lucente}}]{Arina:2017sng}%
  \BibitemOpen
  \bibfield  {author} {\bibinfo {author} {\bibfnamefont {Chiara}\ \bibnamefont
  {Arina}}, \bibinfo {author} {\bibfnamefont {Mihailo}\ \bibnamefont
  {Backovic}}, \bibinfo {author} {\bibfnamefont {Jan}\ \bibnamefont {Heisig}},
  \ and\ \bibinfo {author} {\bibfnamefont {Michele}\ \bibnamefont {Lucente}},\
  }\bibfield  {title} {\enquote {\bibinfo {title} {{Solar $\gamma$-rays as a
  Complementary Probe of Dark Matter}},}\ }\href@noop {} {\  (\bibinfo {year}
  {2017})},\ \Eprint {http://arxiv.org/abs/1703.08087} {arXiv:1703.08087
  [astro-ph.HE]} \BibitemShut {NoStop}%
\end{thebibliography}%

\end{document}